\documentclass[11pt]{article}
\pdfoutput=1
\usepackage{jheppub}
\usepackage{tabu}
\usepackage[vcentermath]{youngtab}
\usepackage[usenames,dvipsnames,table]{xcolor}
\usepackage{graphicx,amsmath,amssymb,amsthm,multirow,array,bm,bbm,esint}
\usepackage[mathscr]{eucal}
\usepackage[bbgreekl]{mathbbol}
\usepackage{epsf,amsfonts}
\usepackage{slashed}
\usepackage[numbers,sort&compress]{natbib}
\usepackage{minibox}
\usepackage{rotating}
\usepackage{pdflscape}
\usepackage{array,tikz-cd}
\usepackage{subcaption}
\usepackage{framed}
\usepackage{mathtools}
\usepackage{tikz} 
\usepackage[compat=1.1.0]{tikz-feynman}
\usepackage{feynmf}
\usepackage{float}
\usepackage{verbatim}

\usepackage{slashed}

\def \Tr{\mbox{Tr\,}}

\def \tr{\mbox{tr\,}}

\newcommand{\ads}{\text{\tiny AdS} }

\newcommand{\im}{\text{Im}}

\newcommand{\const}{\text{\tiny BTZ} }
\newcommand{\constb}{{\tiny\beta} }

\newcommand{\kc}{k_{\tiny cs} }	

\newcommand{\ons}{\text{\tiny on-shell} }	

\newcommand{\ie}{\emph{i.e.}, }
\newcommand{\eg}{\emph{e.g.}, }

\newcommand{\etc}{\emph{etc}.}

\makeatletter
\newcommand{\doubletilde}[1]{{%
  \mathpalette\double@tilde{#1}%
}}
\newcommand{\double@tilde}[2]{%
  \sbox\z@{$\m@th#1\tilde{#2}$}%
  \ht\z@=.9\ht\z@
  \tilde{\box\z@}%
}
\makeatother

\title{Chaos in Three-dimensional Higher Spin Gravity}

\author[a]{Prithvi Narayan}
\author[b]{\!, Junggi Yoon}
\affiliation[\,a]{Department of Physics, Indian Institute of Technology, Palakkad 678557, India}
\affiliation[\,b]{School of Physics, Korea Institute for Advanced Study\\
85 Hoegiro Dongdaemun-gu, Seoul 02455, Republic of Korea.}
\emailAdd{prithvi.narayan@gmail.com}
\emailAdd{junggiyoon@kias.re.kr}

\preprint{{\raggedleft \tt KIAS-P19017 \par} }

\abstract{We discuss $SL(N,\mathbb{C})$ Chern-Simons higher spin gauge theories in Euclidean AdS$_3$. With appropriate boundary term, we derive the higher spin generalization of Schwarzian on-shell action. We investigate gravitationally dressed bi-locals, and we study the soft higher spin mode expansion to obtain soft mode eigenfunctions. We also derive the spin-$s$ eigenfunction from $\mathcal{W}$-Ward identity and a recursion relation. Using the on-shell action, we evaluate the contributions of the soft higher spin modes to the out-of-time-order correlators, and the corresponding Lyapunov exponent of $SL(N)$ higher spin gravity is found to be ${2\pi \over \beta}(N-1)$.
}

\begin{document}
\maketitle

\section{Introduction}
\label{sec:introduction}

Recently, chaotic systems have been extensively investigated in the context of AdS/CFT correspondence. One way to characterize the chaos is the butterfly effect measuring the sensitivity to the initial condition of a system. In classical system, the butterfly effect is measured by ${\delta q(t) \over \delta q(0)}$ which grows exponentially in time in the chaotic system. In quantum theories, the out-of-time-ordered correlator~(OTOC) has emerged as a new way to quantify the quantum chaos~\cite{Shenker:2013pqa,Roberts:2014isa,Shenker:2014cwa,Shenker:2013yza}. The exponential growth rate $\lambda_L$ of OTOCs, so-called Lyapunov exponent, is one of the important measurement of chaos, and it can be considered as a  universal characteristic of chaotic system. Unlike classical system, it was shown~\cite{Maldacena:2015waa} that the Lyapunov exponent of a reasonable\footnote{\eg a system with unitarity and causality \etc} physical system is bounded by ${2\pi \over \beta}$ where $\beta$ is the inverse temperature of the system. \ie
\begin{equation}
    \lambda_L\;\leqq \;{2\pi \over \beta}
\end{equation}
The bound on chaos provides a natural concept of maximally chaotic system. Even before this result was known, it was already anticipated that black holes are maximally chaotic~(the fastest scrambler)~\cite{Sekino:2008he}, and it was indeed shown that the black hole has maximal Lyapunov exponent by various techniques including geodesic distance~\cite{Shenker:2013pqa}, elastic eikonal approximation~\cite{Shenker:2014cwa,Jahnke:2019}, vacuum conformal block~\cite{Fitzpatrick:2016thx,Chen:2016cms,Perlmutter:2016pkf} and the on-shell action~\cite{Maldacena:2016upp,Narayan:2017qtw,Yoon:2017nig,Narayan:2017hvh,Gaikwad:2018dfc,Poojary:2018esz}.

In the context of AdS/CFT correspondence, field theory dual to black hole should also be maximally chaotic. Although many interesting models are believed to be maximally chaotic, it is, in general, difficult to evaluate the OTOCs explicitly. Recently, quantum mechanical models such as Sachdev-Ye-Kitaev~(SYK) models~\cite{Sachdev:1992fk,Polchinski:2016xgd,Jevicki:2016bwu,Maldacena:2016hyu,Jevicki:2016ito,Gross:2016kjj,Fu:2016vas} and SYK-like tensor models~\cite{Gurau:2010ba,Carrozza:2015adg,Witten:2016iux,Gurau:2016lzk,Klebanov:2016xxf} were shown to be maximally chaotic, and they have provide fruitful playground to investigate the black hole physics holographically. See reviews~\cite{Sarosi:2017ykf,Jahnke:2018off} and recent interesting development in this direction~\cite{Das:2017pif,Chaudhuri:2017vrv,Yoon:2017gut,Peng:2017spg,Bulycheva:2017uqj,Krishnan:2016bvg,Krishnan:2017ztz,deMelloKoch:2017bvv,Qaemmaqami:2017jxz,Azeyanagi:2017drg,Bulycheva:2017ilt,Krishnan:2017lra,Kitaev:2017awl,Das:2017wae,Diaz:2018xzt,Giombi:2018qgp,Maldacena:2018lmt,Chang:2018sve,Berkooz:2018qkz,Benedetti:2018goh,Nosaka:2018iat,Peng:2018zap,Ahn:2018sgn,Blommaert:2018oro,Mertens:2018fds,Sun:2019mms,Garcia-Garcia:2019poj,deMelloKoch:2019ywq,Kim:2019upg,Ferrari:2019ogc}.

The low energy dynamics of these maximally chaotic quantum mechanical models is universally governed by Schwarzian action, which plays an important role in the saturation of the chaos bound~\cite{Maldacena:2016hyu}. This Schwarzian action has also been observed as an on-shell action of the dual 2D gravity~\cite{Maldacena:2016upp,Jensen:2016pah,Bagrets:2016cdf,Grumiller:2017qao,Mandal:2017thl} as well as the dimensional reduction of higher dimensional black hole~\cite{Cvetic:2016eiv,Castro:2018ffi,Gaikwad:2018dfc}. Furthermore, it was found as an on-shell action of 3D gravity: metric-like formulation~\cite{Poojary:2018esz} and frame-like Chern-Simons formulation~\cite{Cotler:2018zff}.

In Chern-Simons gravity~\cite{Achucarro:1987vz,Witten:1988hc} with boundary, it is natural to have the Schwarzian on-shell action. The variational principle of Chern-Simons theory with boundary requires an appropriate boundary term. When this boundary term breaks the gauge symmetry on the boundary which corresponds, roughly speaking, to the conformal symmetry on the boundary, one gets anomaly of the conformal symmetry as an on-shell action.

The Chern-Simons gravity can be generalized into Chern-Simons higher spin gravity~\cite{Blencowe:1988gj,Campoleoni:2010zq,Campoleoni:2011hg}. The higher spin gravity~\cite{Fronsdal:1978rb,Fang:1978wz,Fronsdal:1978vb,Vasiliev:1990en,Vasiliev:1995dn,Prokushkin:1998bq,Prokushkin:1998vn,Bekaert:2005vh} in AdS and its dual CFT~\cite{Gaberdiel:2010pz} have provided a promising laboratory to test holography~\cite{Klebanov:2002ja,Sezgin:2002rt, Giombi:2009wh,Papadodimas:2011pf,Hijano:2013fja,Koch:2014aqa,Koch:2014mxa,Jevicki:2015sla}. The quantum chaos in the higher spin theories was explored by $\mathcal{W}$ vacuum blocks~\cite{Perlmutter:2016pkf} and by Wilson line in the charge shock wave background~\cite{David:2017eno}. In the Chern-Simons higher spin gravity, the corresponding higher spin on-shell action will be the anomaly of its asymptotic symmetry, so-called $\mathcal{W}$ symmetry. The form of anomaly, which is the generalization of the Schwarzian derivative, has been observed in the various literature~\cite{Bershadsky:1989mf,Marshakov:1989ca,Li:2015osa,Gonzalez:2018enk,Ma:2019gxy}.

In this paper, we will show how this anomaly can appear as the on-shell action\footnote{See \cite{Gonzalez:2018enk} for the on-shell action for 2D higher spin BF theory.} of the higher spin gravity by choosing a appropriate boundary term and the corresponding boundary condition in~\cite{deBoer:2013gz}. For this, we consider a constant background like BTZ black hole and its smooth fluctuations which does not change the holonomy of the connection up to conjugation. In addition, we study gravitationally dressed bi-locals such as (boundary-to-boundary) Wilson line or master field in the Vasiliev equation, and we investigate the ``soft'' mode expansion of them which leads to the $\mathcal{W}$-Ward identity of bi-locals. Here, ``soft'' denotes a smooth infinitesimal residual gauge transformation from the constant background so that it is not involved with any large gauge transformation. Using the soft mode expansion, we will analyze the quantum chaos of the higher spin gravity based on the on-shell action.

The outline of this paper is as follows. {\bf In Section~\ref{sec: higher spin gravity}}, we review the Chern-Simons higher spin gravity with boundary, and we derive the generalized Schwarzian action as the on-shell action. {\bf In Section~\ref{sec: zero mode eigenfunction}}, we consider master field in Vasiliev equation and Wilson line in the non-constant background, and we study the soft mode expansion thereof. In addition, based on the null relation in $\mathcal{W}_N$ minimal model we derive $\mathcal{W}$-Ward identity. We also construct a recursion relation of the soft mode eigenfunction for general $s$. {\bf In Section~\ref{sec: lyapunov exponent}}, we evaluate the contribution of higher spin fields to the OTOCs to read off Lyapunov exponent of higher spin gravity. {\bf In Section~\ref{sec:conclusion}}, we make concluding remarks and present the future directions. {\bf In Appendix~\ref{app: infinitesimal transf}}, we propose a conjecture on the null relation for general spin $s$ generators in $\mathcal{W}_N$ minimal model. {\bf In Appendix~\ref{app: recursion}}, we provide elaborate on the recursion relations of soft mode eigenfunctions. {\bf In Appendix~\ref{app: lyapunov exponent}}, we estimate the contribution of spin $s$ soft mode to Lyapunov exponent. {\bf In Appendix~\ref{app: 2d gravity}}, we quickly apply our techniques to 2D higher spin BF theory to evaluate the Lyapunov exponent. {\bf In Appendix~\ref{app: Toda Theory}}, we discuss the relation between charges in $SL(3)$ Chern-Simons higher spin gravity and the charge in Toda field theory.

\section{$\mathcal{W}$-Schwarzian Action in Higher Spin Gravity}
\label{sec: higher spin gravity}

\subsection{Review: Chern-Simons Higher Spin Gravity for Euclidean AdS$_3$}
\label{sec: review}

We begin with the review of $SL(N,\mathbb{C})$ Chern-Simons higher spin gauge gravity for Euclidean AdS$_3$~\cite{Blencowe:1988gj,Campoleoni:2010zq,Gutperle:2011kf}. Especially, we will use $(\tau,\bar{\tau})$ formalism in~\cite{deBoer:2013gz}. Also, since we are interested in evaluation of the out-of-time-ordered correlator~(OTOC) from the on-shell action via analytic continuation, we consider Euclidean AdS$_3$ (EAdS$_3$).

The three-dimensional pure higher spin gauge theories in $EAdS_3$ can be formulated by Chern-Simons action with complex Lie algebra $SL(N,\mathbb{C})$:
\begin{equation}
	I_{CS}={i \kc\over 4\pi }\int_{\mathcal{M}} \Tr \left[ CS(A)-CS(\bar{A})\right]\label{def: cs action}
\end{equation}
where $\kc={l\over 4G}$ is Chern-Simons level and $CS(A)$ is defined by
\begin{equation}
	   CS(A)= A\wedge dA +{2\over 3} A\wedge A \wedge A
\end{equation}
In addition, the trace $\Tr$ is defined in term of the trace $\tr$ over the fundamental representation by
\begin{equation}
    \Tr(M)={1\over \tr(L_0L_0)}\tr (M)
\end{equation}
where $L_0, L_{\pm 1}$ are the generator of $sl(2)$ subalgebra of $sl(N)$, and we will consider the principal embedding of $sl(2)$ into $sl(N)$ in this work. Furthermore, we use convention for $sl(N)$ generators such that
\begin{equation}
    (L^{(s)}_n)^\dag=(-1)^n L_{-n}^{(s)} \hspace{8mm} (s=2,3,\cdots, N)
\end{equation}
For the representation for the generators, see Appendix~\ref{app: convention} for $N=2,3$ cases\footnote{For $sl(N)$, see \cite{Hijano:2013fja}. Note that we use the different realization of the $sl(3)$ generators from~\cite{deBoer:2013gz}.}. Also, in our convention, the Chern-Simons connection $A$ is related to its conjugation $\bar{A}$ by
\begin{equation}
	\bar{A}=-A^\dag
\end{equation}
We use the coordinates $(r,z,\bar{z})$ given by
\begin{equation}
    z\equiv\phi+ i {t_E\over l}\hspace{3mm},\hspace{5mm} \bar{z}\equiv \phi- i {t_E\over l}
\end{equation}
The manifold of Chern-Simons gravity for Euclidean AdS$_3$ is a solid torus of which the modular parameter $\tau$ of the boundary torus gives the periodicity of $z$ coordinate:
\begin{equation}
    z\sim z+2\pi \sim z+\tau
\end{equation}
For BTZ black hole, the modular parameter $\tau$ is given by\footnote{Note that $r_--=i r_{\text{\tiny E}}$ is pure imaginary in Euclidean signature so that we will distinguish them clearly.}
\begin{align}
	\tau=& w + {i\beta \over l}= {2\pi i l ( r_+ -i r_{\text{\tiny E}})\over r_+^2-r_-^2}= {2\pi i l\over r_+ + r_-}\cr
	\bar{\tau}=& w - {i\beta \over l}= -{2\pi i l ( r_+ +i r_{\text{\tiny E}})\over r_+^2-r_-^2} =-{2\pi i l\over r_+ - r_-}
\end{align}

One can fix a gauge~\cite{Campoleoni:2010zq,Campoleoni:2011hg} such that
\begin{align}
    A=&b^{-1} (d+ a_z dz + a_{\bar{z}}d\bar{z} ) b\\
	\bar{A}=&b (d+ \bar{a}_z dz + \bar{a}_{\bar{z}}d\bar{z}) b^{-1}
\end{align}
where $b(r)$ is defined by
\begin{equation}
	b=e^{rL_0}
\end{equation}
In Chern-Simons gravity theories, the asymptotic AdS$_3$ condition with a flat boundary metric, which is analogous to the Brown-Henneaux asymptotic AdS boundary condition in the metric-like formulation, is~\cite{Campoleoni:2010zq,Campoleoni:2011hg}
\begin{align}
    \left.A-A_{\ads}\right|_{\partial\mathcal{M}}\sim \phi(1)\label{eq: asymptotic ads condition}
\end{align}
where $A_{\ads}$ is the exact $AdS_3$ solution~\cite{Campoleoni:2010zq,Campoleoni:2011hg}:
\begin{equation}
    A_{\ads}=b^{-1}[\;d + (L_1 + {1\over 4}L_{-1})dz\;] b\hspace{4mm},\hspace{6mm} \bar{A}_{\ads}=b[\;d+ (L_{-1} + {1\over 4}L_1)d\bar{z}\; ]b^{-1}
\end{equation}
The asymptotic AdS$_3$ condition~\eqref{eq: asymptotic ads condition} leads to
\begin{equation}
    a=L_1dz +\cdots\hspace{5mm},\hspace{5mm} \bar{a}=L_{-1}d\bar{z} +\cdots
\end{equation}

The variation of the Chern-Simons action without any boundary term\footnote{We will choose appropriate boundary term soon in~\eqref{eq:Boundary Term in action}}~\eqref{def: cs action} gives
\begin{equation}
    \delta I_{CS}= -{i\kc \over 4\pi } \int_{\partial\mathcal{M} } \Tr \left[ A \wedge \delta A- \bar{A} \wedge \delta \bar{A} \right]
\end{equation}
Hence, we can choose a boundary condition~$A_{\bar{z}}=\bar{A}_z=0$ for consistent variational principle. Furthermore, by using residual gauge symmetry, we can fix the highest weight gauge~\cite{Campoleoni:2010zq,Campoleoni:2011hg}. For $N=3$, the highest weight gauge is given by
\begin{align}
	a_z(z)=& L_1 - {2\pi \over \kc} \mathcal{L}(z) L_{-1} +{\pi \over 2\kc } \mathcal{W}(z) W_{-2}=\begin{pmatrix}
	0 & \sqrt{2}{2\pi  \over \kc }\mathcal{L}(z) & {2\pi \over \kc}\mathcal{W}(z) \\
	\sqrt{2} & 0  & \sqrt{2} {2\pi  \over \kc }\mathcal{L}(z)  \\
	0 & \sqrt{2} & 0\\
	\end{pmatrix}\cr
	\bar{a}_{\bar{z}}(\bar{z})=& L_{-1} - {2 \pi \over \kc}\bar{\mathcal{L}}(\bar{z}) L_{1}  + {\pi \over 2\kc } \bar{\mathcal{W}} (\bar{z}) W_2  =\begin{pmatrix}
	0 & - \sqrt{2} & 0\\
	-\sqrt{2} { 2\pi  \over \kc }\bar{\mathcal{L}} (\bar{z})  & 0 & - \sqrt{2} \\
	-{2\pi \over \kc }\bar{\mathcal{W}}(\bar{z}) & -  \sqrt{2}  { 2\pi  \over \kc }\bar{\mathcal{L}}(\bar{z}) & 0 \\
	\end{pmatrix}
\end{align}

To study generic higher spin black hole, we need to consider the variation of the modular parameter $\tau$ as well as chemical potentials for higher spin charges~\cite{Gutperle:2011kf,Ammon:2011nk,Castro:2011fm,Gaberdiel:2012yb,deBoer:2013gz}. We will use the $(\tau,\bar{\tau})$ formalism in~\cite{deBoer:2013gz} to incorporate the variation of the modular parameters. Although the formalism works for generic chemical potentials, for simplicity, we turn off the chemical potential for higher spin charges - if present they become a source of the spin-$s$ Schwarzian. Hence, in this work, we focus only on BTZ black hole embedded in the higher spin gravity in which the derivation of on-shell action will formally be parallel to that of $SL(2,\mathbb{C})$ case~\cite{Jahnke:2019}.

Recall that the modular parameter appears as the periodicity of the coordinates $(z,\bar{z})$. Hence, for the variation of the modular parameter, it is useful to fix periodicity. For this, we introduce a new coordinate $(w,\bar{w})$ defined by
\begin{equation}
	z={1-i {\tau\over 2\pi} \over 2} w + {1+ i {\tau\over 2\pi}\over 2 } \bar{w}
\end{equation}
and, the coordinates $w$ and $\bar{w}$ has a fixed periodicity:
\begin{equation}
    w\sim w+2\pi \sim w+2\pi i
\end{equation}
By fixing the periodicity of $w,\bar{w}$, the boundary metric and the boundary volume element depends on the modular parameter:
\begin{align}
	ds^2=& dzd\bar{z} = \left|\left({1-i{\tau\over 2\pi}\over 2}\right)dw +\left({1+i{\tau\over 2\pi} \over 2 }\right)d\bar{w} \right|^2\\
	i dw \wedge d\bar{w}=& {4\pi dz^2 \over \im (\tau)}
\end{align}
where $dz^2={i\over 2} dz\wedge d\bar{z}$. It was the key point in ~\cite{deBoer:2013gz} to keep the boundary volume element $w\wedge \bar{w}$ under the variation of the action. In the variation of the Chern-Simons bulk action~\eqref{def: cs action}, it is convenient to use $(w, \bar{w})$ coordinates because the differential form $dw$ and $d\bar{w}$ are not varied under the variation. Then, after varying the action, one can return to $(z,\bar{z})$ coordinates where the variation of $a_w, a_{\bar{w}}$ leads to the variation of the modular parameter $\tau,\bar{\tau}$ because the linear transformation from $(a_w,a_{\bar{w}})$ to $(a_z,a_{\bar{z}})$ depends on $\tau$ and $\bar{\tau}$:
\begin{equation}
    a_w=\left({1-i{\tau\over 2\pi} \over 2 }\right) a_z +\left({1-i{\bar{\tau}\over 2\pi} \over 2 }\right) a_{\bar{z}}
\end{equation}
With the boundary term chosen in~\cite{deBoer:2013gz}
\begin{equation}\label{eq:Boundary Term in action}
	I_b=-{\kc \over 2\pi } \int_{\partial \mathcal{M} } d^2 z\; \Tr \left(  (a_z- 2 L_1)a_{\bar{z}} \right) -{\kc \over 2\pi } \int_{\partial \mathcal{M}} d^2 z\; \Tr \left(  (\bar{a}_{\bar{z}}- 2 L_{-1})\bar{a}_{z} \right) \ ,
\end{equation}
the variation of the total action $I_{\text{\tiny tot}}\equiv I_{CS}+I_b$ is 
\begin{align}
	\delta I_{\text{\tiny tot}} =&- i \kc \int_{\partial \mathcal{M} } {d^2z\over 2\pi \im (\tau) } \Tr \left[ (a_z-L_1)\delta\left( (\bar{\tau}-\tau)a_{\bar{z}}\right)+\left({a_z^2\over 2}+ a_za_{\bar{z}} -{\bar{a}_z^2\over 2} \right)\delta \tau \right.\cr
	&\hspace{35mm}\left.-(-\bar{a}_{\bar{z}}-L_{-1})\delta\left( (\bar{\tau}-\tau)\bar{a}_{z}\right)+\left({\bar{a}_{\bar{z}}^2\over 2}+ \bar{a}_{\bar{z}}\bar{a}_{z} -{a_{\bar{z}}^2\over 2} \right)\delta \bar{\tau} \right]
\end{align}
For consistent boundary condition, we impose impose a boundary condition 
\begin{align}
	a_{\bar{z}}= \bar{a}_z=0\qquad,\qquad \delta\tau=\delta\bar{\tau}=0\label{eq: boundary condition}
\end{align}
Note that since we turn off the chemical potential for higher spin charges, the variation of the total action is simpler than that of~\cite{deBoer:2013gz}. Then, using the boundary condition we obtain the on-shell action of $I_{\text{\tiny tot}}$:
\begin{align}
	I_\ons =  {i \kc\over 2\pi} \int {d^2 z\over \im (\tau) }  \Tr\left[ {\tau\over 2}a_z^2  - {\bar{\tau}\over 2}\bar{a}^2_{\bar{z}}   \right]\label{eq: onshell action deboer}
\end{align}
Note that though \cite{deBoer:2013gz} mainly analyzed a constant solution $a$ and $\bar{a}$ with chemical potential, they already pointed out that their formulation can be applied to the non-constant solutions.

\subsection{Asymptotic AdS Solution and On-shell Action}
\label{sec: asymptotic ads solution}

In this section, we will work out the asymptotic AdS solution and the corresponding on-shell action for the case of $SL(3,\mathbb{C})$ explicitly. Due to the boundary condition in \eqref{eq: boundary condition}, by using residual gauge symmetry, we can fix gauge connection to be 
\begin{align}
	a_z(z)=& L_1 - {2\pi \over \kc} \mathcal{L}(z) L_{-1} +{\pi \over 2\kc } \mathcal{W}(z) =\begin{pmatrix}
	0 & \sqrt{2}{2\pi  \over \kc }\mathcal{L}(z) & {2\pi \over \kc}\mathcal{W}(z) \\
	\sqrt{2} & 0  & \sqrt{2} {2\pi  \over \kc }\mathcal{L}(z)  \\
	0 & \sqrt{2} & 0\\
	\end{pmatrix}\cr
	\bar{a}_{\bar{z}}(\bar{z})=& L_{-1} - {2 \pi \over \kc}\bar{\mathcal{L}}(\bar{z}) L_{1}  + {\pi \over 2\kc } \bar{\mathcal{W}} (\bar{z})   =\begin{pmatrix}
	0 & - \sqrt{2} & 0\\
	-\sqrt{2} { 2\pi  \over \kc }\bar{\mathcal{L}} (\bar{z})  & 0 & - \sqrt{2} \\
	-{2\pi \over \kc }\bar{\mathcal{W}}(\bar{z}) & -  \sqrt{2}  { 2\pi  \over \kc }\bar{\mathcal{L}}(\bar{z}) & 0 \\
	\end{pmatrix}\label{eq: gauge condition}
\end{align}
Then, the on-shell action in \eqref{eq: onshell action deboer} can be written as
%
%
\begin{align}
	I_\ons=  i \int {d^2 z\over \im (\tau) }  \left[ \tau \mathcal{L}(z) - \bar{\tau}  \bar{\mathcal{L}}(\bar{z}) \right]\label{eq: onshell action}
\end{align}
In this work, we will study non-constant gauge connection $a$ and $\bar{a}$ that are connected to a fixed constant solution $a_{\const}$ (\eg BTZ black hole embedded in the higher spin gravity) by smooth gauge transformation. \ie
\begin{equation}
	a= h^{-1} a_{\const}h+ h^{-1}d h 
\end{equation}
where smooth gauge transformation means that $h(z)$ is a holomorphic residual gauge transformation parameter which can be smoothly connected to identity. We are interested in the simplest constant connection $a_\const$ without spin-3 charge:
\begin{equation}
	a_\const=\begin{pmatrix}
	0 &  \sqrt{2} {2\pi \over \kc}\mathcal{L}_0& 0\\
	\sqrt{2} & 0 &   \sqrt{2}  {2\pi \over \kc}\mathcal{L}_0  \\
	0 & \sqrt{2} & 0\\
	\end{pmatrix}
\end{equation}
where $\mathcal{L}_0$ is a constant. Under the smooth residual gauge transformation, the holonomy is not changed up to similarity transformation.
\begin{equation}
	\text{Hol}_{\mathcal{C}}(A)\equiv\mathcal{P}\exp[-\int_{\mathcal{C}} A]=b^{-1}h^{-1} e^{-\mathbb{w}} h b 
\end{equation}
Note that a holonomy along the contractible cycle of the solid torus should be trivial, namely, it belong to the center of the $SL(N)$ gauge group~\cite{Gutperle:2011kf,Castro:2011fm,Castro:2011iw}. The Euclidean time circle $\mathcal{C}$ is the contractible on in the BTZ black hole, and the smoothness of its holonomy leads to
\begin{equation}
	\mathbb{w}=\tau a_z +\bar{\tau} a_{\bar{z}}=u^{-1} ( 2\pi i L_0) u
\end{equation}
for some matrix $u$. From this condition, one can find the relation between the constant $\mathcal{L}_0$, constant energy-momentum tensor of the BTZ black hole, and the modular parameter $\tau$:
\begin{align}
	\tr (\mathbb{w}^2)=& \tr \left[\begin{pmatrix}
	0 & -{ 2\pi \tau \mathcal{L}_0 \over \kc } \sqrt{2} & 0 \\
	\sqrt{2}\tau   & 0 & -{ 2\pi \tau \mathcal{L}_0 \over \kc } \sqrt{2} \\
	0 & \sqrt{2}\tau & 0\\
	\end{pmatrix}^2\right]={ 16\pi \tau^2  \mathcal{L}_0 \over \kc }  = -8\pi^2  
\end{align}
and, we have
\begin{align}
	\tau =i\pi \sqrt{\kc \over 2\pi \mathcal{L}_0 }\hspace{4mm},\hspace{6mm}\bar{\tau} = -i \pi \sqrt{\kc \over 2\pi \bar{\mathcal{L}}_0 }\label{eq: result of smoothness of holonomy}
\end{align}
Recall that we fix the modular parameter under the variation of the action. This corresponds to fixing the constant $\mathcal{L}_0$ and $\bar{\mathcal{L}}_0$. Therefore, our on-shell action captures the (fixed) BTZ black hole geometry embedded in higher spin gravity with $\tau,\bar{\tau}$ together with the smooth fluctuation thereof.

Now, we will calculate $\mathcal{L}(z)$ in the non-constant connection $a(z)$ in terms of the smooth residual gauge transformation parameter $h(z)$ which transforms the constant background $a_\const$ to $a(z)$. For arbitrary constant $\mathcal{L}_0$, this can be worked out in principle as follow. Using the Gauss decomposition of the gauge parameter $h(z)$, one can find a generic residual gauge symmetry parameter which keep the gauge condition that we chose in~\eqref{eq: gauge condition}. However, among the generic residual gauge symmetry parameter, it is not easy to distinguish smooth one from large gauge transformation which will change the holonomy. Unlike $SL(2)$ case where we know the correct answer for any $\mathcal{L}_0$, for now we do not have a guiding principle to derive the form of $\mathcal{L}(z)$ or $\mathcal{W}(z)$ for the non-zero $\mathcal{L}_0$, and this corresponds to find ``the finite temperature $SL(N)$ Schwarzian derivative''.\footnote{For ``zero temperature Schwarzian derivative'', see Appendix~\ref{app: Toda Theory} and \cite{Bershadsky:1989mf,Marshakov:1989ca,Gonzalez:2018enk,Ma:2019gxy}. Note that \cite{Gonzalez:2018enk} proposed a conjecture on the finite temperature $SL(3)$ Schwarzian.}

At least, an infinitesimal residual gauge transformation is smooth, and one can obtain $\mathcal{L}(z)$ and $\mathcal{W}(z)$ perturbatively. But, unlike $SL(2)$ case, it is not easy to integrate the perturbative expression of $\mathcal{L}(z)$ and $\mathcal{W}(z)$ to obtain the full higher spin Schwarzian derivative at finite temperature, either. Nevertheless, for our purpose, it is enough to evaluate $\mathcal{L}(z)$ perturbatively because the leading contribution of boundary higher spin modes to OTOCs can be evaluated by quadratic on-shell action. Hence, let us consider an infinitesimal residual gauge transformation by
\begin{equation}
	h= \mathbb{I}+\epsilon\; \lambda^{(1)} +{1\over 2}\epsilon^2\left[ (\lambda^{(1)})^2 + \lambda^{(2)}\right]+\cdots\label{eq: infinitesimal residual transf parametrization}
\end{equation}
where $\lambda^{(1)}, \lambda^{(2)}\in sl(3)$ are parametrized by
\begin{align}
	\lambda^{(1)}= \begin{pmatrix}
	\varphi_1^{(1)} & e_{-1}^{(1)} & e_{-3}^{(1)} \\
	e_1^{(1)}  &  -\varphi_1^{(1)} +\varphi_2^{(1)} & e_{-2}^{(1)}  \\
	e_3^{(1)} & e_2^{(1)}  &  -\varphi_2^{(1)} \\
	\end{pmatrix}\\
	\lambda^{(2)}= \begin{pmatrix}
	\varphi_1^{(2)} & e_{-1}^{(2)} & e_{-3}^{(2)} \\
	e_1^{(2)}  &  -\varphi_1^{(2)} +\varphi_2^{(2)} & e_{-2}^{(2)}  \\
	e_3^{(2)} & e_2^{(2)}  &  -\varphi_2^{(2)} \\
	\end{pmatrix}
\end{align}
The infinitesimal gauge transformation leads to the expansion of the on-shell action in~\eqref{eq: onshell action}:
\begin{equation}
    I_{\ons}=I_{\ons}^{(0)}+\epsilon\; I_{\ons}^{(1)}+\epsilon^2\; I_{\ons}^{(2)}+\cdots\label{eq: expansion of on-shell action}
\end{equation}
We demand that the gauge transformation of the constant connection $a_\const$ by $h(z)$ keeps the gauge condition in~\eqref{eq: gauge condition} order by order. \ie
\begin{align}
	a=&h^{-1} a_\const h + h^{-1} \partial_z h \cr
	=&  a_\const + \epsilon \begin{pmatrix}
	0 & \sqrt{2} {2\pi \over \kc} \mathcal{L}^{(1)}& {2\pi \over \kc}\mathcal{W}^{(1)}\\
	0 & 0 & \sqrt{2}{2\pi \over \kc} \mathcal{L}^{(1)}  \\
	0 & 0 & 0\\
	\end{pmatrix} + \epsilon^2 \begin{pmatrix}
	0 & \sqrt{2}{2\pi \over \kc}\mathcal{L}^{(2)} & {2\pi \over \kc}\mathcal{W}^{(2)}\\
	0 & 0 & \sqrt{2}{2\pi \over \kc}\mathcal{L}^{(2)} \\
	0 & 0 & 0\\
	\end{pmatrix}+\cdots \label{eq:Quadratic Expansion of a under h}
\end{align}
In the first order, one can determine $\varphi_1^{(1)},\varphi_2^{(1)},e_2^{(1)},e_{-1}^{(1)},e_{-2}^{(1)},e_{-3}^{(1)}$ as a linear combinations of $e_1^{(1)}$  and $e_3^{(1)}$. Also, by demanding the gauge condition in the second order, one can express $\varphi_1^{(2)},\varphi_2^{(2)},e_2^{(2)},e_{-1}^{(2)},e_{-2}^{(2)},e_{-3}^{(2)}$ in terms of the quadratics of $e_1^{(1)}$  and $e_3^{(1)}$ and the linear combinations of $e_1^{(2)}$  and $e_3^{(2)}$. As a result, the first order $\mathcal{L}^{(1)}$ and $\mathcal{W}^{(1)}$ gives the $\mathcal{W}_3$ algebra as asymptotic symmetry with identification $c=6\kc={3l\over 2G}$~\cite{Campoleoni:2010zq,Henneaux:2010xg,Campoleoni:2011hg}. By redefining\footnote{Here, we absorbed the infinitesimal parameter $\varepsilon$ into $\eta(z)$ and $\zeta(z)$}
\begin{align}
	\epsilon\; e_1^{(1)}(z)=&\eta(z)-{1\over 2\sqrt{2} } \zeta'(z) \label{def: fz}\\
	\epsilon\; e_3^{(2)}(z)=&\zeta(z)\label{def: gz}
\end{align}
one can express the second order $\mathcal{L}^{(2)}$ and $\mathcal{W}^{(2)}$ in terms of $\eta(z)$ and $\zeta(z)$:
\begin{align}\label{eq:schwarzzian for T}
	\epsilon^2\mathcal{L}^{(2)}=& {\kc\over 16\pi }\left(  -\left({2\pi \over \tau}\right)^2 [\eta']^2 + [\eta'']^2 \right)+ {\kc \over 384\pi }\left( 4\left({2\pi \over \tau}\right)^4[\zeta']^2- 5 \left({2\pi \over \tau}\right)^2 [\zeta'']^2+ [\zeta''']^3 \right)\\
	\label{eq:schwarzzian for W}
	\epsilon^2\mathcal{W}^{(2)}=&{\kc\over 24\sqrt{2}\pi } \left( 4\left({2\pi \over \tau}\right)^4 \eta' \zeta'- 5 \left({2\pi \over \tau}\right)^2\eta'' \zeta''+ \eta''' \zeta'''\right)
\end{align}
up to total derivatives. Here, we used the relation between the modular parameter $\tau$ and $\mathcal{L}_0$ in \eqref{eq: result of smoothness of holonomy} derived from the smoothness of holonomy along the contractible cycle. Expanding $\eta(z)$ and $\zeta(z)$ by
\begin{align}
	\eta(z)=\sum_{n} \eta_n\; e^{-{2\pi i n z \over \tau } }\hspace{4mm},\hspace{6mm} \zeta(z)=\sum_{n} \zeta_n\; e^{-{2\pi i n z \over \tau } }\label{eq: mode expansion of soft mode}
\end{align}
the quadratic on-shell action $I^{(2)}_\ons$ in \eqref{eq: expansion of on-shell action} becomes
\begin{align}
	\epsilon^2 \; I_\ons^{(2)}=&   {2\pi^4 i \kc \over \tau^3} \sum_{n\geqq 2} n^2(n^2-1) \eta_{-n}\eta_n + {\pi^6 i \kc \over 3\tau^5}\sum_{n \geqq 3} n^2(n^2-1)(n^2-4) \zeta_{-n} \zeta_n\cr
	& - (\text{anti-holomorphic})
\end{align}
Note that the on-shell action for $\eta_n$ and $\zeta_n$ vanishes for $n=0,\pm1$ and $n=0,\pm1,\pm2$, respectively. They are the $SL(3,\mathbb{C})$ isometry of the constant connections. In the higher spin black hole, $SL(3,\mathbb{C})$ isometry of AdS vacuum is supposed to be broken because of the periodicity of $\varphi$. But, because we are working in the covering space of $\varphi$, we still have $SL(3,\mathbb{C})$ isometry. Now, one can read off the two point function of boundary soft modes: 
\begin{alignat}{3}
	&\langle \eta_{-n}\eta_n\rangle=&& {\kappa_2\over n^2(n^2-1)} \hspace{10mm}&(|n|\geqq 2)\label{eq: soft mode 2pt s2}\\
	&\langle \zeta_{-n}\zeta_n\rangle=&& {\kappa_3\over n^2(n^2-1)(n^2-4)}\hspace{5mm}&(|n|\geqq 3)\label{eq: soft mode 2pt s3} \\
\end{alignat}
where the coefficient $\kappa_2$ and $\kappa_3$ are given by
\begin{align}
	\kappa_2\equiv { \tau^3\over 2\pi^4 i \kc } \qquad,\qquad \kappa_3\equiv {3\tau^5\over \pi^6 i \kc }
\end{align}

\section{Gravitational Dressing and Soft Higher Spin Expansion}
\label{sec: zero mode eigenfunction}

One way to diagnose quantum chaos is by the out-of-time-ordered correlators of matter field or equivalently by scattering of shock waves in the dual bulk theory. However, the three-dimensional pure higher spin gauge theory does not contain a matter field. This problem is more pronounced in Chern-Simons formulation, since being topological, it is not easy to couple the higher spin fields to matter which is not topological in AdS$_3$. The construction of an action of interacting higher spin gauge theories with matter is also a challenging problem.

Although it is difficult to construct interaction with matter field, one may consider a scalar field in the probe limit. And, in a simple background geometry such as BTZ black hole embedded in higher spin gravity, the boundary-to-boundary propagator of the probe scalar field is nothing but two point function in CFT$_2$ with suitable conformal dimension. However, the evaluation of OTOCs from the on-shell action via analytic continuation requires the $\mathcal{W}$ transformation of the two point function, at least, infinitesimally. In contrast to the conformal transformation, the $\mathcal{W}$ transformation of two point function is not well-understood due to the non-linearity of $\mathcal{W}_N$ algebra.

Despite of the above obstacles, we could find ways to deal with those difficulties. For the evaluation of OTOCs via analytic continuation, we first need to calculate Euclidean four point function on the boundary. For this, we consider a particular type of four point function which can be viewed as two point function of bi-local operators $\Phi(z_1,\bar{z}_1;z_2,\bar{z}_2)$ and its leading contribution to the four point function (equivalently, two point function of two bi-locals) is the product of one-point function of each bi-locals. 
\begin{equation}
	\langle \Phi_1(z_1,\bar{z}_1;z_2,\bar{z}_2)\Phi_2(z_3,\bar{z}_3;z_4,\bar{z}_4)  \rangle=G_1(z_{12},\bar{z}_{12})G_2(z_{34},\bar{z}_{34})+\cdots
\end{equation}
where $G_i(z,\bar{z})$ $(i=1,2)$ is the one-point function of the bi-local operator corresponding to a boundary-to-boundary two point function:
\begin{equation}
    G_i(z_{12},\bar{z}_{12})\equiv \langle \Phi_i(z_1,\bar{z}_1;z_2,\bar{z}_2)\rangle\hspace{8mm} (i=1,2)
\end{equation}
We will consider a gravitationally dressed bi-local field, namely, the bi-local field in the non-constant background. In particular, for a non-constant background which is connected to a constant background (\eg BTZ black hole) by infinitesimal gauge transformation, one can expand the gravitationally dressed bi-local field with respect to the soft modes. 
\begin{equation}
    \Phi^{\text{\tiny dressed}}(z_1,\bar{z}_1;z_2,\bar{z}_2)=G(z_1,\bar{z}_1;z_2,\bar{z}_2)+ \epsilon\; G^{(1)}(z_1,\bar{z}_1;z_2,\bar{z}_2)+\cdots
\end{equation}
Recall that an infinitesimal residual gauge transformation induces an infinitesimal $\mathcal{W}$ transformation on the boundary. Therefore, the soft mode expansion of the gravitationally dressed bi-local field, which is generated by an infinitesimal residual gauge transformation from constant connection to non-constant one, gives the $\mathcal{W}$ transformation of the leading term, namely, $\mathcal{W}$ transformation of the boundary two point function.

In~\cite{Dobrev:1977qv,deMelloKoch:2018ivk}, the complete set of conformal partial wave functions has been constructed by correlation function of bi-local operator and spin-$s$ primary operator. In this spirit, one can also induce the soft mode eigenfunction by taking correlation function between the dressed bi-local operator and the soft higher spin mode: 
\begin{equation}
    \langle \epsilon\; \Phi^{\text{\tiny dressed}}(z_1,\bar{z}_1;z_2,\bar{z}_2)\rangle \sim G^{(1)}(z_1,\bar{z}_1;z_2,\bar{z}_2) +\cdots\label{eq: overlap soft mode and bilocals}
\end{equation}
Recall that the soft mode expansion of the dressed bi-local operator defines $G^{(1)}$ \ie $\delta_\epsilon \Phi^{\text{\tiny dressed}}= G^{(1)}$. Therefore, \eqref{eq: overlap soft mode and bilocals} can be viewed as $\mathcal{W}_3$-Ward identity.

In the next few sections, we will discuss the form of $G^{(1)}$ using various arguments. First, we will consider the master field of matter in the Vasiliev equation in Section~\ref{sec: vasiliev equation} and a Wilson line in Section~\ref{sec: wilson line} as a gravitationally dressed bi-local operator, and we will study the soft mode expansion thereof. In addition, from the null relation of primary operators in higher spin AdS$_3$/CFT$_2$ duality to find the $\mathcal{W}$-Ward identity for its two point function in Section~\ref{sec: ward identity}. We also discuss recursion relations of the soft mode eigenfunctions in Section~\ref{sec: recursion relation}.

\subsection{Matter Master Field in Vasiliev Equation}
\label{sec: vasiliev equation}

Let us consider the $\mathcal{W}_N$ minimal model at semi-classical limit where 'tHooft coupling constant $\lambda$ is taken to be $-N$ in the higher spin AdS$_3$/CFT$_2$~\cite{Gaberdiel:2012ku,Perlmutter:2012ds,Hijano:2013fja}. In the semi-classical limit, the gauge sector becomes $sl(N)\times sl(N)$ Chern-Simons gravity where the gauge connection $A$ is $sl(N)$ matrix of which equation of motion is given by
\begin{equation}
	dA+A\wedge A=0
\end{equation}
In addition, the propagating scalar field can be described by master field $C$ which is a $N\times N$ matrix~\cite{Hijano:2013fja}: 
\begin{align}
	dC+A C- C\overline{A}=0
\end{align}
Note that the physical scalar field corresponds to the trace of the master field $C$, and other components can be expressed in terms of its derivatives~\cite{Ammon:2011ua,Hijano:2013fja}. Note that the equations of motion are invariant under the gauge transformation
\begin{align}
	\delta A=& d \xi + A\xi -\xi A\\
	\delta C=& -\xi C + C \bar{\xi}
\end{align}
As in the previous section, we take into account a constant background and its fluctuation connected by smooth residual gauge transformation:
\begin{equation}
    A_z=b^{-1}(h^{-1} a_\const h + h^{-1} \partial_z h) b\hspace{4mm},\hspace{6mm} A_r=b^{-1}\partial_r b
\end{equation}
where $	b(r)\equiv e^{r L_0}$.

The solution of the equation of motion for $C$ in the constant background is easily found to be~\cite{Hijano:2013fja}
\begin{equation}
	C(r,z,\bar{z})=b^{-1}(r) e^{-a_\const z} c_0e^{\bar{a}_\const \bar{z} } b^{-1}(r)\label{eq: master field soluion1}
\end{equation}
where $c_0$ is a constant matrix. It was proven~\cite{Hijano:2013fja} that $\tr (C)$ satisfies the Klein-Gordon equation for any choice of $c_0$. In particular, for $\mathcal{L}_0=\bar{\mathcal{L}}_0=0$, it satisfies
\begin{equation}
    [\partial_r^2 +2\partial_r +4e^{-2r}\partial\bar{\partial}-(N^2-1)]\tr(C)=0
\end{equation}
Note that with the choice of ``the highest weight state'' $c_0$
\begin{equation}
	(c_0)_{ij} =\delta_{i1} \delta_{j1}\;\; ,
\end{equation} 
it was also shown that $\tr(C)$ gives the boundary-to-bulk propagator~\cite{Hijano:2013fja}.

One can also study the master field in a non-stationary background. The solution with non-constant connection can be written as Wilson line: 
\begin{align}
	&C(r_1,z_1,\bar{z}_1;z_2,\bar{z}_2) \cr
	=& \lim_{r_2\rightarrow \infty}  e^{2hr_2} b^{-1}(r_1)\mathcal{P}\exp\left[-\int^{z_1}_{z_2} a   \right] b(r_2) \tilde{c}_0 b(r_2) \mathcal{P}\exp\left[-\int^{\bar{z}_2}_{\bar{z}_1} \bar{a}  \right]  b^{-1}(r_1)
\end{align}
where $\tilde{c}_0$ is a constant matrix. Here, one need to choose a reference point $z_2,r_2$ corresponding to the other point of Wilson line, and if $r_2 \rightarrow \infty$, one can interpret the $z_2$ as the position of source on the boundary. We took limit $r_2\rightarrow \infty $ with factor $e^{2hr_2}$ where the conformal dimension $h=\bar{h}$ of scalar field is given by
\begin{equation}
h=\bar{h}=-{N-1\over 2}
\end{equation}
The negative conformal dimension reflects that the semi-classical limit of higher spin AdS$_3$/CFT$_2$ is non-unitary~\cite{Gaberdiel:2012ku,Hijano:2013fja}. Furthermore, according to the dictionary of the higher spin AdS$_3$/CFT$_2$ \cite{Gaberdiel:2010pz,Gaberdiel:2011zw}, the scalar field in the $hs[\lambda]$ higher spin gravity has conformal dimension $h={1+\lambda\over2}$. In the semi-classical limit where we perform the analytic continuation $\lambda\rightarrow-N$, the conformal dimension of the scalar field becomes 
\begin{equation}
    h={1+\lambda\over 2}\qquad\Longrightarrow \qquad h=-{N-1\over 2}
\end{equation}
This agrees with the conformal dimension of the master field.

In this prescription with Wilson line for the master field, it is natural to have ``the highest weight state'' due to the regularization. \ie
\begin{equation}
    c_0\equiv \lim_{r_2\rightarrow \infty}  e^{2hr_2}b(r_2) \tilde{c}_0 b(r_2)\qquad\Longrightarrow \qquad (c_0)_{ij}\sim \delta_{i1}\delta_{j1}
\end{equation}
Furthermore, $\tr(C)$, represents the bulk-to-boundary propagator, and we also take limit $r_1 \rightarrow \infty$ in $\tr(C)$, which leads to boundary-to-boundary propagator in the non-constant background. We take it as our bi-local field:
\begin{equation}
	\Phi^{\text{\tiny dressed}}_{\text{\tiny master}}(z_1,\bar{z}_1;z_2,\bar{z}_2)\equiv\lim_{r \rightarrow \infty } e^{2h r_1}\tr [C(r_1,z_1,\bar{z}_1;z_2,\bar{z}_2)]\label{def: dressed master field}
\end{equation}
Recall that one can consider the soft mode expansion of the non-constant connection around a constant one. And, the boundary-to-boundary propagator in the non-constant background can be understood as the gravitationally dressed master field. Using the residual gauge transformation, we can express the dressed master field as follow.
\begin{align}\label{eq:dressed c}
	&C^{\text{\tiny dressed}}(r_1,z_1,\bar{z}_1;z_2,\bar{z}_2)\cr
	=&  b^{-1}(r_1) h^{-1}(z_1)e^{- a_\const (z_1-z_2)} h(z_2) c_0\bar{h}(\bar{z}_2) e^{- \bar{a}_\const(\bar{z}_2-\bar{z}_1)} \bar{h}^{-1}(\bar{z}_1) b^{-1}(r_1)
\end{align}
For the soft mode expansion
\begin{equation}\label{eq:Def Boundary to Boundary Propagator}
	\Phi^{\text{\tiny dressed}}_{\text{\tiny master}}(z_1,\bar{z}_1;z_2,\bar{z}_2)=G(z_1,\bar{z}_1;z_2,\bar{z}_2)+G^{(1)}(z_1,\bar{z}_1;z_2,\bar{z}_2)+\cdots\ ,
\end{equation}
we use the infinitesimal residual gauge transformation in~\eqref{eq: infinitesimal residual transf parametrization} parametrized by $\eta(z)$ and $\zeta(z)$ (See \eqref{def: fz} and \eqref{def: gz}).  The leading bi-local field $G(z_1,\bar{z}_1;z_2,\bar{z}_2)$ is a boundary-to-boundary propagator in the constant background:
\begin{equation}
       G(z_1,\bar{z}_1;z_2,\bar{z}_2)=\left[{{\pi^2\over \tau\bar{\tau} } \over \sin \left({\pi z_{12} \over\tau }\right) \sin \left({\pi \bar{z}_{12} \over\bar{\tau} }\right)}\right]^{2h}
\end{equation}
For the sub-leading one, using the mode expansion of $\eta(z)$ and $\zeta(z)$ in~\eqref{eq: mode expansion of soft mode}, we have 
\begin{equation}
	{G^{(1)}(z_1,\bar{z}_1;z_2,\bar{z}_2)\over G(z_1,\bar{z}_1;z_2,\bar{z}_2)}=\sum_{|n|\geqq2} \eta_n f_{2,n}(z_1,z_2)+ \sum_{|n|\geqq 3}\zeta_n f_{3,n}(z_1,z_2)+ (\text{anti-holomorphic})
\end{equation}
where soft mode eigenfunction $f_{s,n}(z_1,z_2)$ ($s=2,3$) is found to be  
\begin{align}
    f_{2,n}\equiv&\gamma_2\; e^{-{2\pi  i n \chi \over \tau}  } \left[n\cos { 2\pi n \sigma \over \tau } -{\sin { 2\pi n\sigma\over \tau}\over \tan {2\pi \sigma\over \tau }}\right]\label{def: f2n }\\
    f_{3,n}\equiv&\gamma_3\; e^{-{2\pi  i n \chi \over \tau}  } \left[ 2 n^2 \sin { 2\pi n \sigma \over \tau } + 6 n  { \cos { 2\pi n \sigma \over \tau }\over \tan { 2\pi \sigma \over \tau }} - 2{ 1+2\cos^2 { 2\pi  \sigma \over \tau }\over\sin^2{ 2\pi \sigma \over \tau }} \sin { 2\pi n \sigma \over \tau }\right]\label{def: f3n}
\end{align}
where we defined $(\chi,\sigma)$ in terms of the bi-local coordinates $(z_1,z_2)$ by
\begin{equation}
    \chi={1\over 2}(z_1+z_2)\hspace{4mm},\hspace{6mm}\sigma={1\over 2}(z_1-z_2)
\end{equation}
and the coefficient $\gamma_s$ ($s=2,3$) is found to be
\begin{equation}
    \gamma_2\equiv - {4 \pi i h \over \tau} \hspace{4mm},\hspace{6mm} \gamma_3\equiv  {4 \pi^2 i h(2h+1) \over \tau^2}\label{eq: gamma coefficient}
\end{equation}
Also, we used the relation~\eqref{eq: result of smoothness of holonomy} derived from the smoothness of holonomy:
\begin{equation}
    \mathcal{L}_0= -{\kc \pi \over 2 \tau^2}\hspace{4mm},\hspace{6mm} \bar{\mathcal{L}}_0= -{\kc \pi \over 2 \bar{\tau}^2}
\end{equation}

For propagators in the BTZ background, we need to impose the periodicity of the angular coordinate $\phi\sim \phi+2\pi$, which can be realized by the method of images. This is also involved with the non-contractible cycle of BTZ black hole. We will discuss this issue in the next section together with Wilson line.

\vspace{3mm}

\noindent
{\bf Remarks on the highest weight prescription:}\;\; Before ending this section, we make a remark on the choice of ``the highest weight state'' $(c_0)_{ij}\sim \delta_{i1}\delta_{j1}$ from the point of view of the residual gauge transformation. From the solution in~\eqref{eq: master field soluion1} of the master field in the constant background, the residual gauge transformation of the connection $a$ and $\bar{a}$ induces the transformation of $c_0$ where the information about the boundary operator $(z_2,\bar{z}_2)$ is encoded. To see this, note that the residual gauge transformation by $h$ will transform the $c_0$ as follow.
\begin{equation}
    c_0\qquad\longrightarrow \qquad h(z_2) c_0 \bar{h}(\bar{z}_2)
\end{equation}
For simplicity, let us consider the action of holomorphic gauge transformation, and we work out explicitly for $sl(2)$ case with $\mathcal{L}_0=0$:
\begin{align}
	&(\mathbb{I}+\epsilon\; \lambda^{(1)}(z))\begin{pmatrix}
	(c_0)_{11} & (c_0)_{12}\\
	(c_0)_{21} & (c_0)_{22}\\
	\end{pmatrix}=c_0 +(c_0)_{11}\begin{pmatrix}
	-{1\over 2} f'(z)& 0\\
	f(z) & 0\\
	\end{pmatrix} +(c_0)_{12}\begin{pmatrix}
	0& -{1\over 2} f'(z)\\
	0 & f(z) \\
	\end{pmatrix}\cr
	&\hspace{20mm}+(c_0)_{21}\begin{pmatrix}
	-{1\over 2} f''(z) & 0\\
	{1\over 2}f'(z) & 0\\
	\end{pmatrix}+(c_0)_{22}\begin{pmatrix}
	0 & -{1\over 2} f''(z)\\
	0 & {1\over 2}f'(z) \\
	\end{pmatrix}\label{eq: transf of boundary}
\end{align}
where we parametrized the infinitesimal residual gauge parameter by
\begin{equation}
    \lambda^{(1)}(z)=\begin{pmatrix}
    -{1\over 2}f'(z) & -{1\over 2} f''(z)\\
    f(z) & {1\over 2} f'(z)\\
    \end{pmatrix}
\end{equation}
Recall the conformal transformation of primary $ \phi$  and its descendant $L_{-1}\phi=\partial_z \phi$ by infinitesimal variation $\epsilon(z)$ is given by:
\begin{align}\label{eq:Transformation of Matter Field}
	\delta_\epsilon \phi=& -\epsilon(z) \partial_z\phi- h \partial_z \epsilon(z) \phi\\
	\delta_\epsilon \partial_z \phi=& - h\partial_z^2 \epsilon(z) \phi - (h+1) \partial_z \epsilon(z) \partial_z\phi-\epsilon(z) \partial_z^2\phi\label{eq: conformal transf of descendant}
\end{align} 
where $h$ is the conformal dimension of $\phi$. Comparing \eqref{eq: transf of boundary} with \eqref{eq: conformal transf of descendant}, the holomorphic part of $(c_0)_{11}, (c_0)_{12}$ modes transform like primaries with $h=-{1\over 2}$ while $(c_0)_{21}$ and $(c_0)_{22}$ components behave like descendants. Also, note that the last term in \eqref{eq: conformal transf of descendant} does not appear in the \eqref{eq: transf of boundary}. This would be because the operator dual to the matter field in the semi-classical limit has the conformal dimension $h=-{1\over 2}$ and $\partial^2\phi$ is a null state for $SL(2)$ case (See Section~\ref{sec: ward identity}). In the same way with anti-holomorphic transformation, one can show that $(c_0)_{11}$ and $(c_0)_{21}$ modes are primaries while $(c_0)_{12}$ and $(c_0)_{22}$ are descendants. Hence, $(c_0)_{11}$ mode is the primary operator in both holomorphic and anti-holomorphic sector, and this is why $(c_0)_{11}$ mode gives the correct bulk-to-boundary correlator. In Section~\ref{sec: ward identity}, we will discuss how a special class of operators (\eg the primary operator dual to scalar field in higher spin gravity) behave under $\mathcal{W}$ transformation. Based on this observation, one can see that ``the highest state'' mode indeed transforms as a primary for the case of $N=3$.

\subsection{Wilson Line}
\label{sec: wilson line}

The Wilson line captures interesting physics in gauge theories. In the Chern-Simons gravity as a gauge theory, the Wilson line plays an important role. In~\cite{Ammon:2013hba,deBoer:2013vca,Castro:2014tta,deBoer:2014sna,Castro:2018srf}, the Wilson line has been investigated in the Chern-Simons (higher spin) gravity to define entanglement entropy holographically. This proposal is very natural because Wilson line between two boundary points gives geodesic distance of them, which is equivalent to the Ryu-Takayanagi prescription~\cite{Ryu:2006bv,Ryu:2006ef}. In 2D, the Wilson line was studied in the Schwarzian theory to evaluate the OTOCs~\cite{Blommaert:2018oro,Blommaert:2018iqz,Blommaert:2019hjr}. In previous section, the Wilson line also appears as a solution of the equation of motion of the master field.

For Wilson line in Chern-Simons gravity, for fixed end points, one can still consider various Wilson line objects depending on the choice of states of the Wilson line operator.\footnote{For instance, as in previous section one can take the trace of Wilson line operator. Or, we can also take a particular component. In fact, the solution of master field can be viewed as the highest weight state for $(z_2,\bar{z}_2)$ when we remove the trace by exchanging the holomorphic and anti-holomorphic part. } Those were extensively studied from the point of view of Ishibashi states in the dual CFT$_2$~\cite{Castro:2018srf}. In this paper, we will simply consider the following Wilson line~\cite{deBoer:2013vca} between $(r_1,z_1,\bar{z}_1)$ and $(r_2,z_2,\bar{z}_2)$ as our bi-local field:
\begin{equation}
        \Phi_{\text{\tiny Wilson}}(z_1,\bar{z}_1;z_2,\bar{z}_2)=\lim_{r\rightarrow \infty} e^{4h r}\tr\left[\mathcal{P}\exp\left(-\int^{r,z_1}_{r,z_2} A \right)\mathcal{P}\exp\left(-\int_{r,\bar{z}_1}^{r,\bar{z}_2} \bar{A} \right)\right]\label{def: wilson line}
\end{equation}
where we regularize Wilson line from boundary to boundary symmetrically. Then, this is equivalent to the definition of the gravitationally dressed master field in~\eqref{def: dressed master field} with $(c_0)_{ij}=\delta_{ij}$. Hence, under the soft mode expansion, we have the same result.
\begin{equation}
	\Phi^{\text{\tiny dressed}}_{\text{\tiny Wilson}}(z_1,\bar{z}_1;z_2,\bar{z}_2)=G(z_1,\bar{z}_1;z_2,\bar{z}_2)+G^{(1)}(z_1,\bar{z}_1;z_2,\bar{z}_2)+\cdots\ ,
\end{equation}
where the leading and the sub-leading bi-locals are given by
\begin{align}
       G(z_1,\bar{z}_1;z_2,\bar{z}_2)=&\left[{{\pi^2\over \tau\bar{\tau} } \over \sin \left({\pi z_{12} \over\tau }\right) \sin \left({\pi \bar{z}_{12} \over\bar{\tau} }\right)}\right]^{2h}\\
       {G^{(1)}(z_1,\bar{z}_1;z_2,\bar{z}_2)\over G(z_1,\bar{z}_1;z_2,\bar{z}_2)}=&\sum_n \eta_n f_{2,n}(z_1,z_2)+ \zeta_n f_{3,n}(z_1,z_2)+ (\text{anti-holomorphic})
\end{align}
with the same $f_{2,n}$ and $f_{3,n}$ as in the master field as before in \eqref{def: f2n } and \eqref{def: f3n}. Though the mathematical definition of the simple Wilson line in our consideration is the same as that of the master field solution, the Wilson line can give geometric insights. First of all, the Wilson line will give the geodesic distance between two boundary points. In 3D, the geodesic corresponds to the Ryu-Takanayagi surface, and the Wilson line can provide the entanglement entropy~\cite{Ammon:2013hba,deBoer:2013vca}. Following~\cite{deBoer:2013vca}, we define the holographic entanglement entropy (on the constant Euclidean time slice) by
\begin{equation}
	S_{EE}(\Delta\phi)\equiv -{\kc\over 2h}\log\left. e^{4 h r_0}\Phi^{\text{\tiny dressed}}_{\text{\tiny Wilson}}(z_1,\bar{z}_1;z_2,\bar{z}_2)\right|_{t_E=0, \phi=\phi_1-\phi_2}\label{def: ee wilson line}
\end{equation}
where $h$ is the conformal dimension of the operator at the end points of Wilson line. For our case, it is given by
\begin{equation}
	h=-{1\over 2}(N-1)
\end{equation}
Also, we retrieved the UV divergence by the radial cutoff $r_0$ (See \eqref{def: wilson line}). As pointed out in~\cite{deBoer:2013vca}, this reproduces known entanglement entropy in the various 3D background. In particular, we are interested in BTZ background where the non-contractible cycle along $\phi$. For simplicity, we will consider the non-rotating BTZ black hole\footnote{See~\cite{Jahnke:2019} for rotating BTZ black hole in $SL(2)$ Chern-Simons gravity.}, and the Wilson line via \eqref{def: ee wilson line} indeed reproduces the entanglement entropy in the BTZ background~\cite{deBoer:2013vca}
\begin{equation}
	S_{EE}(\Delta\phi)=  {c\over 6} \log \left[ { \pi^2 \beta^2  \over l^2 e^{-2r_0}} \sinh^2 \left({ \pi l \Delta \phi  \over \beta }\right) \right]
\end{equation}
Here, we used the relation between the central charge $c$ and Chern-Simons level $\kc$
\begin{equation}
	c=6\kc
\end{equation}
and we also used the modular parameter of the non-rotating black hole:
\begin{equation}
	\tau= {2\pi i l\over r_+}={i \beta \over l}\hspace{3mm},\hspace{6mm} \bar{\tau}=-{2\pi i l\over r_+}=-{i \beta \over l}
\end{equation}
The existence of the non-contractible cycle could make Wilson line ambiguous. But, by considering the covering space of $\phi$, one can incorporate the contribution of horizon into Wilson line. First, let us consider $\phi=2\pi$ where one end of the Wilson line rotate around whole boundary and comes back to the other point. in high temperature $\beta\gg l$, one can also recover the Bekenstein-Hawking formula for the BTZ black hole
\begin{equation}
	S_{EE}(2\pi)= {\pi r_+\over 4G}+\cdots
\end{equation}
where we used $\kc={l\over 4G}$. As the Wilson line winds the horizon further, it is natural to encode the winding number into bi-local fields like Wilson line as well as master field.  
\begin{equation}
        \Phi_{m}^{\text{\tiny dressed}}(z_1,\bar{z}_1;z_2,\bar{z}_2)\equiv \Phi^{\text{\tiny dressed}}(z_1+2\pi m,\bar{z}_1+2\pi m ;z_2,\bar{z}_2)\hspace{10mm}(m\in \mathbb{Z})\label{def: image of dressed bilocal}
\end{equation}
For AdS vacuum where the modular parameter is given by $2\pi$,  the holonomy around the $\phi$ cycle is trivial. \ie $\Phi_{m}^{\text{\tiny dressed}}(z_1,\bar{z}_1;z_2,\bar{z}_2)= \Phi^{\text{\tiny dressed}}(z_1,\bar{z}_1;z_2,\bar{z}_2)$, but it is not true for BTZ black hole. For non-constant background, one can also take the soft mode expansion of $\Phi_{m}^{\text{\tiny dressed}}$:
\begin{equation}
	\Phi_{m}^{\text{\tiny dressed}}(z_1,\bar{z}_1;z_2,\bar{z}_2)=G_m(z_1,\bar{z}_1;z_2,\bar{z}_2)+ G_m^{(1)}(z_1,\bar{z}_1;z_2,\bar{z}_2)+\cdots
\end{equation}
When we evaluate the boundary-to-boundary propagator $G_{\text{\tiny BTZ}}$ in the BTZ black hole background, we impose the periodicity in $\phi$ by hand by summing up those wound bi-locals~\cite{KeskiVakkuri:1998nw,Kraus:2002iv}
\begin{equation}
	G_{\text{\tiny BTZ}}(z_1,\bar{z}_1;z_2,\bar{z}_2)=\sum_m G_m(z_1,\bar{z}_1;z_2,\bar{z}_2)=\sum_{m\in \mathbb{Z} } {1\over [\sin{\pi (z_{12} + 2\pi m )\over \tau } \sin {\pi (\bar{z}_{12} + 2\pi m )\over \bar{\tau} } ]^{2h}  }
\end{equation} 
Now, we will study the sub-leading term of $\Phi_{m}^{\text{\tiny dressed}}$ in the soft mode expansion:
\begin{align}
       G^{(1)}_m(z_1,\bar{z}_1;z_2,\bar{z}_2)=&\left[\sum_{|n|\geqq 2} \eta_n \tilde{f}_{2,n;m}(z_1,z_2)+ \sum_{|n| \geqq 3}\zeta_n \tilde{f}_{3,n;m}(z_1,z_2)\right]G_m(z_1,\bar{z}_1;z_2,\bar{z}_2)\cr
       &+ (\text{anti-holomorphic})
\end{align}
where $\tilde{f}_{2,n;m}(z_1,z_2)$ and $\tilde{f}_{3,n;m}(z_1,z_2)$ is found to be
\begin{align}
    \tilde{f}_{2,n;m}\equiv& \gamma_2\;e^{-{2\pi  i n \chi \over \tau}  } \left[n\cos { 2\pi n \sigma \over \tau } -{\sin { 2\pi n\sigma\over \tau}\over \tan {2\pi (\sigma+ \pi m)\over \tau }}\right]\\
    \tilde{f}_{3,n;m}\equiv& \gamma_3\;e^{-{2\pi  i n \chi \over \tau}  } \left[ 2 n^2 \sin { 2\pi n \sigma \over \tau } + 6 n  { \cos { 2\pi n \sigma \over \tau }\over \tan { 2\pi (\sigma+ \pi m) \over \tau }} - 2{ 1+2\cos^2 { 2\pi  (\sigma+ \pi m) \over \tau }\over\sin^2{ 2\pi (\sigma+ \pi m) \over \tau }} \sin { 2\pi n \sigma \over \tau }\right]
\end{align}
where $\gamma_2$ and $\gamma_3$ is given in \eqref{eq: gamma coefficient}. Note that for $m\ne 0$ we have
\begin{align}
	\tilde{f}_{2,0,m}(\chi,\sigma)=\tilde{f}_{3,0,m}(\chi,\sigma)=0
\end{align}
while
\begin{align}
	\tilde{f}_{2,\pm,m}(\chi,\sigma)\ne0 \quad,\quad \tilde{f}_{3,\pm1,m}(\chi,\sigma)\ne 0\quad,\quad \tilde{f}_{3,\pm2,m}(\chi,\sigma)\ne 0
\end{align}
This reflects the well-known fact that in BTZ black hole the global $sl(N)$ isometry is broken to its Cartan $[U(1)]^{\oplus N-1}$.

\subsection{Null Relation and $\mathcal{W}$-symmetry Ward Identity}
\label{sec: ward identity}

In the last two sub-section, we have investigated the bi-local fields in the non-constant background and the soft mode expansion thereof. As we discussed, the soft mode, which generates (infinitesimal) residual gauge transformation, will induce the $\mathcal{W}$-transformations on the boundary. Hence, the soft mode expansion of the gravitationally dressed bi-local field give the $\mathcal{W}$-Ward identity on the boundary. In CFT, we know how two point function of any primary operators transforms under the conformal symmetry, and we have seen that soft mode expansion indeed reproduce the conformal transformation of two point function~\cite{Maldacena:2016hyu,Maldacena:2016upp,Narayan:2017qtw,Yoon:2017nig,Narayan:2017hvh,Gaikwad:2018dfc,Poojary:2018esz}. In the previous section, we obtained soft mode expansion for $\mathcal{W}_3$ symmetry, which is related to the $\mathcal{W}_3$ transformation of two point function in CFT with $\mathcal{W}_3$ symmetry. However, from the point of view of CFT, this is, in general, difficult to study due to the non-linearity of $\mathcal{W}$-algebra, and a general formula is not known at least to author's knowledge.

Though the $\mathcal{W}_3$ transformation of a generic primary would be difficult, one might be able to find it for a special case where non-linearity is suppressed. In this section, we derive the $\mathcal{W}_s$-transformation ($s=3,4$) of a special primary based on its null relation in the 'tHooft limit of the $W_N$ minimal model~\cite{Gaberdiel:2011zw,Chang:2011vka,Jevicki:2013kma,Chang:2013izp}. Also, in Appendix~\ref{app: infinitesimal transf} we make a conjecture to extend the null relations and we derive the soft mode eigenfunctions for general spin $s$.

In $\mathcal{W}_N$ minimal model, a primary operator can be labelled by two Young tableaux $(\Lambda_+;\Lambda_-)$~\cite{Gaberdiel:2010pz,Gaberdiel:2011zw}. In particular, a primary operator $\phi_1\equiv ({\tiny \yng(1)};0)$ is dual to the scalar field in the dual higher spin gravity. It was shown~\cite{Gaberdiel:2011zw} that because of null relation, the action of generators $W^{(s)}_{-n}$ on a primary operator $\phi_1$ of conformal dimension $h \equiv {1 \over 2} (1+\lambda)$ is proportional to $L_{-1}^n\phi_1$ in large $c$. For instance, the null relation of $\phi_1$ was found to be
\begin{align}
	W^{(3)}_{0} \phi_1=&w^{(3)} \phi_1\label{eq: op w action1} \hspace{20mm} w^{(3)}=-(1+\lambda)(2+\lambda) \\
	W^{(3)}_{-1}\phi_1=& {3w^{(3)}\over 2h} L_{-1} \phi_1\\
	W^{(3)}_{-2}\phi_1=& {3w^{(3)}\over h(2h+1)} L_{-1}^2 \phi_1 
\end{align}
And, for its conjugate $\bar{\phi}_1\equiv ({\tiny \overline{\yng(1)}};0)$, we have
\begin{align}
	W^{(3)}_{0} \bar{\phi}_1=&-w^{(3)} \bar{\phi}_1\\
	W^{(3)}_{-1}\bar{\phi}_1=& -{3w^{(3)}\over 2h} L_{-1} \bar{\phi}_1\\
	W^{(3)}_{-2}\bar{\phi}_1=& -{3w^{(3)}\over h(2h+1)}L_{-1}^2 \bar{\phi}_1 \label{eq: op w action2}
\end{align}
%
%
%
Note $\phi_1$ and $\bar{\phi}_1$ have opposite $W_3$ charge. Note that this primary operator corresponds to the master field and Wilson line in the previous section in the semi-classical limit.

At zero temperature, from the mode expansion of $W^{(s)}(z)$ 
\begin{equation}
	W^{(s)}(z)=\sum_{n} z^{-n-s}W^{(s)}_{n}\ ,
\end{equation}
the OPE of $W^{(s)}(z)$ with the primary field $\phi_1(w)$ defined via $W^{(s)}_n \phi_1 = 0, \forall n >0$ gives
\begin{align}
	&-{1\over 2\pi i}\oint dz \;\zeta^{(s)}(z) W^{(s)}(z)\phi_1(w)= -{1\over 2\pi i}\sum_n \oint dz \; {\zeta^{(s)}(z)  \over (z-w)^{n+s} } (W^{(s)}_{n}\phi_1)(w)\cr
	=&-\sum_{m=0}^{s-1} {1\over m!} \partial_z^{m}\zeta^{(s)}(z)(W^{(s)}_{m-s+1}\phi_1)(w)\label{eq: ward identity w3}
\end{align}
Using \eqref{eq: op w action1}$\sim$\eqref{eq: op w action2} the Ward identity for $\mathcal{W}_3$ symmetry can be written as
\begin{align}
	&\delta_\epsilon \langle \phi_1(z_1)\bar{\phi}_1(z_2)\rangle \cr
	=& - w^{(3)}\left[{1\over 2} \partial_1^2 \zeta_1+ {3 \over 2h}\partial_1\zeta_1 \partial_1 + {3 \over h(2h+1)} \zeta_1\partial_1^2 \; - \; (1\rightarrow  2)  \right]  \langle \phi_1(z_1) \bar{\phi}_1(z_2)\rangle 
\end{align}
where $\zeta_i\equiv\zeta(z_i)$ and $\partial_i \equiv {\partial\over \partial z_i}$ $(i=1,2)$. Then, inserting the two point function at zero temperature
\begin{equation}
	\langle \phi_1(z_1)\bar{\phi}_1(z_2)\rangle={1\over (z_1-z_2)^{2h} }
\end{equation}
and using parametrization $\zeta(z) = \sum_n \zeta_n z^{n+2}$, we have the infinitesimal transformation of two point function under $\mathcal{W}_3$ symmetry as:
\begin{align}
	&{\delta_\epsilon \langle \phi_1(z_1)\bar{\phi}_1(z_2)\rangle\over \langle \phi_1(z_1)\bar{\phi}_1(z_2)\rangle}\cr
	=&  w^{(3)}\sum_{n}\zeta_n\left[-{1\over 2} n^2(z_1^n-z_2^n)+{3\over 2}n {z_1+z_2\over z_1-z_2}(z_1^n+z_2^n)-{z_1^2+4z_1z_2+z_2^2\over(z_1- z_2)^2}(z_1^n-z_2^n)\right]\label{eq: zero temperature function}
\end{align}
At finite temperature, one can transform the zero temperature result in~\eqref{eq: zero temperature function} by\footnote{Here, wee use $\tau=2\pi $ for simplicity, and, we will recover it if necessary.}
\begin{equation}
	z\quad\longrightarrow \quad e^{-i z}
\end{equation}
or, one can also use the finite temperature representation of $L_{-1}$ in \eqref{eq: ward identity w3}
\begin{equation}
	L_{-1}= e^{i z}(\partial_z +i h)
\end{equation}
Then, we can obtain
\begin{align}
	&{\delta_\epsilon \langle \phi_1(z_1)\bar{\phi}_1(z_2)\rangle\over  \langle \phi_1(z_1)\bar{\phi}_1(z_2)\rangle}\cr
	=&{i w^{(3)}\over 2} \sum_n \epsilon_n e^{-in \left({z_1+z_2\over 2}\right)} \left[ 2 n^2 \sin{nz_{12}\over 2} + 6 n  { \cos {nz_{12}\over 2}\over \tan {z_{12}\over 2}} - 2{ 1+2\cos^2 {z_{12}\over 2}\over\sin^2{z_{12}\over 2}} \sin{nz_{12}\over 2}\right]
\end{align}
where $\epsilon_i\equiv \epsilon(z_i)$ and $\partial_i=\partial_{z_i}$ $(i=1,2)$. This agrees with the soft mode eigenfunction derived from the soft mode expansion of the dressed bi-locals. In the same way, one can further obtain the transformation of two point function under $W^{(4)}$ symmetry.  The null relations of $\phi_1$ are given by~\cite{Gaberdiel:2011zw}  
\begin{align}
	W^{(4)}_{0} \phi_1=&w^{(4)} \phi_1\\
	W^{(4)}_{-1}\phi_1=& {4w^{(4)}\over 2h} L_{-1} \phi_1\\
	W^{(4)}_{-2}\phi_1=& {10w^{(4)}\over 2h(2h+1)} L_{-1}^2 \phi_1 \\
	W^{(4)}_{-3}\phi_1=& {20w^{(4)}\over 2h(2h+1)(2h+2)} L_{-1}^3 \phi_1 
\end{align}
where $w^{(4)}$ denotes the $W^{(4)}_0$ charge of the operator $\phi_1$. This gives
\begin{align}
	&\delta_{\zeta} \langle \phi_1(z_1)\bar{\phi}_1(z_2)\rangle \cr
	=& - w^{(4)}\left[{1\over 6} \partial_1^3 \zeta_1 + {2 \over 2h}\partial_1^2\zeta_1 \partial_1 + {10 \over 2h(2h+1)} \partial_1 \zeta_1\partial_1^2+ {20 \over 2h(2h+1)(2h+2)}  \zeta_1\partial_1^3 \right.\cr
	&\left.\hspace{15mm}+ (1\rightarrow 2)  \right] \langle \phi_1(z_1)\bar{\phi}_1(z_2)\rangle
\end{align}
And, the transformation of two point function at finite temperature is found to be
\begin{align}
	&{\delta_{\zeta} \langle \phi_1(z_1)\bar{\phi}_1(z_2)\rangle\over \langle \phi_1(z_1)\bar{\phi}_1(z_2)\rangle} =- {w^{(4)}\over 6}\sum_n  \zeta_n e^{-in\left({z_1+z_2\over 2}\right)} \cr
	&\times \left[ 2n^3 \cos {nz_{12}\over 2} -12 n^2 {\sin{nz_{12}\over 2}\over \tan{z_{12}\over 2}} +n(7-15\cot^2{z_{12}\over 2}-15 \csc^2{z_{12}\over 2})\cos{nz_{12}\over 2}  \right.\cr
	&\hspace{5mm} \left. +(5\cot^3{z_{12}\over 2}+25\cot {z_{12}\over 2} \csc^2{z_{12}\over 2}-7\cot {z_{12}\over 2})\sin {nz_{12}\over 2}  \right]
\end{align}
In principle, one can confirm this result from $SL(4,\mathbb{C})$ Chern-Simons gravity although it would be much more tedious calculation. However, from the null relation for $s=3,4$, we could find a pattern of the null relations, and we make a conjecture on null relation for general $s$ in Appendix~\ref{app: infinitesimal transf}. Namely, we conjecture that
\begin{equation}
	W^{(s)}_{-n}\phi_1\sim L_{-1}^s \phi_1
\end{equation}
From this conjecture, we could obtain the soft mode eigenfunction for arbitrary $s$. Surprisingly, this result agrees with another conjecture on the recursion relation of those eigenfunctions. Based on the soft mode eigenfunction for $s=1,2$, \cite{Yoon:2017nig} conjectured a criteria on the recursion relation to construct the eigenfunction for spin $s=n$ from those of $s=1,2,\cdots, n-1$. In Section~\ref{sec: recursion relation}, we complete the conjecture to find a concrete recursion relation to satisfy the criteria in~\cite{Yoon:2017nig}. And, the resulting eigenfunctions perfectly agree with those from the null relation. Furthermore, these eigenfunction satisfies a Casimir differential equation as expected, and we could finally find that those eigenfunctions have been observed as a basis in the large $q$ limit together with $\beta\mathcal{J}\rightarrow \infty $ limit of SYK model~\cite{Maldacena:2016hyu}.

\subsection{Recursion Relation of Soft Mode Eigenfunction}
\label{sec: recursion relation}

In general $SL(N,\mathbb{C})$ Chern-Simon higher spin theory, it is practically difficult to derive the soft mode eigenfunction for arbitrary spin-$s$ explicitly. But, if there is underlying mathematical structure of those eigenfunctions, one might be able to derive them without referring to Chern-Simons higher spin gravity.

We begin with two eigenfunctions for $s=1,2$~\cite{Yoon:2017nig}:
\begin{align}
	f_{1,n}(z_1,z_2)=&{e^{-in \left({z_1+z_2 \over 2}\right)} \over \sin {z_{12}\over 2 }} \sin {nz_{12}\over 2} \\
	f_{2,n}(z_1,z_2)=&{e^{-in \left({z_1+z_2\over 2}\right)} \over \sin {z_{12}\over 2 }}\left[n\cos {nz_{12}\over 2} -{\sin { nz_{12}\over 2}\over \tan{z_{12}\over 2}}\right]		
\end{align}
where we again use holomorphic sector only and choose the (inverse) temperature to be $\tau=2\pi$ for simplicity. Note that we include $\sin {z_{12}\over 2}$ in the denominator which comes from the measure of the $SL(2)$ invariant inner product~\cite{Maldacena:2016hyu,Murugan:2017eto}. It is also convenient to define the center of coordinate $\chi$ and the relative coordinate $\sigma$ by
\begin{equation}
	\chi \equiv{1\over 2} (z_1+z_2)\hspace{5mm},\hspace{5mm} \sigma \equiv {1\over 2}(z_1-z_2)
\end{equation}
By an abuse of notation, we will also redefine $f_{i,n}(\chi+\sigma,\chi-\sigma) \rightarrow f_{i,n}(\chi,\sigma)$ only in this section. We then have a simpler expression
\begin{align}
	f_{1,n}(\chi,\sigma)=&e^{-in \chi}{ \sin { n \sigma}\over \sin \sigma}\\
	f_{2,n}(\chi,\sigma)=&{e^{-in \chi }\over \sin \sigma} \left[n\cos {n\sigma } -{\sin { n\sigma}\over \tan \sigma}\right]	
\end{align}
In the rest of this section we will find the modes $f_{s,n}$ for all $s,n$ by demanding certain conditions. From the translation symmetry of center of coordinate, which is a part of bi-local $SL(2)$ symmetry, we expect 
\begin{equation}\label{eq:Def g}
 f_{s,n}(\chi,\sigma) \equiv e^{-in\chi} g_{s,n}(\sigma)   
\end{equation}
for some $g_{s,n}(\sigma)$.  For our purpose, we need to introduce two ingredients. First one is the $SL(2)$ invariant inner product on the space of functions $f_{s,n}$~\cite{Maldacena:2016hyu,Murugan:2017eto}
\begin{align}
	\langle f_{s,n}, f_{s',n'}\rangle= {1\over 2\pi} \int_0^{2\pi}  d\chi \int_0^\pi d\sigma\; f^\ast_{s,n} (\chi,\sigma)  f_{s',n'}(\chi,\sigma)
\end{align}
Note that we already absorbed the measure into $f_{s,n}$. This inner product can be naturally defined in the SYK model. Due to the translational symmetry of the center of coordinates, the inner product is diagonal in $n$-space. On the $g_{s,n}(z)$ function defined via \eqref{eq:Def g}, this inner product (for the same $n$) translates to 
\begin{equation}\label{eq:inner product on g}
    \langle g_{s,n}, g_{s',n}\rangle=  \int_0^\pi d\sigma\; g^\ast_{s,n} (\sigma)  g_{s',n}(\sigma)
\end{equation}
Another ingredient we need is a recursive way to build the functions $g_{s,n}(\sigma)$. Based on the observation that $f_{2,n}(\chi,\sigma) = \partial_z f_{1,n}(\chi,\sigma)$, it was conjectured in~\cite{Yoon:2017nig} that a mode $f_{s,n} $ is a linear combination of differential operators\footnote{From the point of view of bi-local map in the higher spin AdS/CFT, the relative coordinates $\sigma$ is related to the spin $s$ as well as radial coordinates of AdS.} $\partial_\sigma$ acting on modes $f_{s',n}$ for all $s'<s$. We can make a reasonable conjecture that the set of functions $g_{s,n}(\sigma)$ satisfy the following two conditions:
\begin{align}\label{eq:Zero Mode Condition 1}
	I \hspace{5mm}&:\hspace{5mm} g_{s,n}(\sigma)=\partial_\sigma g_{s-1,n}(\sigma) +\sum_{s'=0}^{s-2} P_{s,s',n}(\partial_\sigma) g_{s',n}(\sigma)\\
	\label{eq:Zero Mode Condition 2}
	II \hspace{5mm}&:\hspace{5mm} 	\langle g_{s,n},g_{s',n} \rangle \sim \delta_{s,s'} 
\end{align}
where $P_{s,s',n}(\partial_\sigma)$ is a polynomial in $\partial_\sigma$. For the given $g_{1,n}(\sigma), g_{2,n}(\sigma)$ it is easy to find explicitly such polynomials $P_{s,s',n}(\partial_\sigma)$ satisfying \eqref{eq:Zero Mode Condition 2} for low spins.  For instance, we found that~\cite{Yoon:2017nig}
\begin{align}
	g_{3,n}(\sigma)=&\partial_z g_{2,n}(\sigma) +{1\over 3} (n^2-1)g_{1,n}(\sigma) \cr
	g_{4,n}(z)=&\partial_z g_{3,n}(\sigma)  + {4\over 15}(n^2-4)g_{s,n}(\sigma)
\end{align}
and so on. Continuing for a few more orders, we could guess that the correct recursion relation is 
\begin{equation}
	g_{s+2,n}=\partial_\sigma g_{s+1,n}+ F_{s,n} g_{s,n}  \label{eq: recursion relation}
\end{equation}
where $F_{s,n}$ is defined by
\begin{equation}
	F_{s,n}\equiv{s^2 (n^2-s^2)\over 4s^2-1} 
\end{equation}
In addition, the initial data for the recursion relation are given by
\begin{align}
	g_{1,n}(\sigma)=&{ \sin { n \sigma}\over \sin \sigma}\\
	g_{2,n}(\sigma)=&{1\over \sin \sigma} \left[n\cos {n\sigma } -{\sin { n\sigma}\over \tan \sigma}\right]	
\end{align}
In Appendix~\ref{app: recursion}, we show that the solution of the recursion relation in~\eqref{eq: recursion relation} is found to be
\begin{equation}
    g_{s,n}(\sigma)  = \begin{cases}\;\; {[(s-1)!]^2 \over (2s-3)!!}   \ { e^{i\pi s } \over \sin \sigma} \left( e^{in\sigma} P_{s-1}^{n,-n}(\sigma) - e^{-in\sigma} P_{s-1}^{-n,n}(\sigma) \right)\quad& \text{for} \;\;|n|\geqq s\\
    \;\;0& \text{for} \;\;|n|< s
    \end{cases}
    \label{eq:g modes not normalized}
\end{equation}
where $P_{n}^{\alpha,\beta}$ is a Jacobi Polynomial. Furthermore, we also show that $f_{s,n}$'s are a solution of a second order differential equation:
\begin{equation}
	\left[-\partial_\chi^2+ \partial_\sigma^2 -{s(s-1)\over \sin^2 \sigma}\right] [ f_{s,n}(\chi,\sigma) \sin \sigma]=0\label{eq: differential eq for fsn}
\end{equation}
Note that this is exactly the same as the differential equation found by~\cite{Maldacena:2016hyu} for the eigenfunction of the four point kernel in the large $q$ limit with $v\rightarrow 1$ of SYK model. Note that the $v\rightarrow 1$ limit corresponds to $\beta\mathcal{J}\rightarrow \infty $ limit in SYK model. In addition, in the zero temperature limit where $\sin \sigma$ is replaced by $\sigma$, such a differential equation is found in~\cite{deMelloKoch:2018ivk} as Casimir differential equation of the bi-local CFTs.

Note the function $f_{s,n}$ diagonalizes the translation generator of center of coordinates. From \eqref{eq: differential eq for fsn}, one can easily show that $f_{s,n}$ are orthogonal. \ie
\begin{equation}
	\langle f_{s,n},f_{s',n'}\rangle=N_{s,n}\delta_{s,s'} \delta_{n,n'}
\end{equation}
where we found (See Appendix~\ref{app: recursion})
\begin{equation}
	N_{s,n}\equiv n\pi (2s-1)\left[{(s-1)!\over (2s-1)!! }\right]^2 \prod_{j=1}^{s-1} (n^2-j^2)
\end{equation}

\section{OTOC and Lyapunov Exponent}
\label{sec: lyapunov exponent}

The Lyapunov exponent diagnosing the quantum chaos is found to be ${2\pi \over \beta}$ for the black hole background in the Einstein gravity. The maximal Lyapunov exponent comes from the graviton exchanges in the high energy scattering. When the scattering can be mediated by higher spin field in addition to graviton, the overall Lyapunov exponent could be changed because one can expect that the Lyapunov exponent by the exchange of the spin $s$ field~\footnote{It was observed in~\cite{Haehl:2018izb} that the ``pole-skipping'' of $W_3$  current leads to the Lyapunov exponent ${4\pi \over \beta}$. } is found to be~\cite{Perlmutter:2016pkf}
\begin{equation}
	\lambda_{L}^{(s)}={2\pi (s-1) \over \beta}
\end{equation}
This seemingly violates the chaos bound. It was shown~\cite{Perlmutter:2016pkf} that in $hs[\lambda]$ higher spin gravity which contains infinite tower of higher spin fields, the exponential growth coming from all higher spin fields forms a geometric series, and the overall Lyapunov exponent becomes zero. Because the $hs[\lambda]$ higher spin theories are unitary for $\lambda\in [0,1]$, the Lyapunov exponent should not exceed ${2\pi \over \beta}$. This is also consistent because the system with infinitely many conserved charge would not be chaotic. On the other hand, the 3D higher spin gravity in the semi-classical limit, the infinite tower of higher spin fields is truncated, and we have only a finite number of higher spin fields. Hence, the geometric series will terminate at a finite spin $s$, and therefore, it will violate the chaos bound. In fact, this is not surprising because this theory is not unitary.

For evaluation of the out-of-time-ordered correlator, we will start from Euclidean four point functions with a particular configuration. As we discussed in Section~\ref{sec: zero mode eigenfunction}, we also consider four point function which can be evaluated by two point function of bi-local field (\eg master field or Wilson line). In the non-constant background, we consider the two point function of the gravitationally dressed bi-local fields, and the soft mode expansion allow us to compute the higher spin contribution to the Euclidean correlator. 
\begin{align}
	F(1,2,3,4)\equiv&\langle\Phi^{\text{\tiny dressed}}(1,2)\Phi^{\text{\tiny dressed}}(3,4) \rangle=G(1,2)G(3,4)+\langle G^{(1)}(1,2)G^{(1)}(3,4)\rangle +\cdots \cr
	&= G(1,2)G(3,4)+ \sum_{s,n} \langle \zeta_n^{(s)} \zeta_{-n}^{(s)} \rangle [\delta_{\zeta_n^{(s)}}G(1,2)]\;[\delta_{\zeta_{-n}^{(s)}}G(3,4)]+\cdots\label{eq: euclidean 4pi for otoc}
\end{align}
where $\zeta^{(s)}_n$ is the spin $s$ soft mode. From the on-shell action, one can read off the two point function of soft modes, which is proportional to ${1\over \kc}\sim {1\over c}$. Therefore, compared to the leading disconnected two point function, the contribution of higher spin field is suppressed by ${1 \over c}$, and this determine the scrambling time to be of order $\log c$.

In BTZ black hole, we also have to take the periodicity of the angle $\phi$ into account. For this, we consider the soft mode expansion of each image of the dressed bi-local field $\Phi^{\text{\tiny dressed}}_{m}(z_1,\bar{z}_1;z_2,\bar{z}_2)$ in~\eqref{def: image of dressed bilocal} due to the periodicity of $\phi\sim \phi+2\pi$, and we denote it by $\tilde{f}_{n,m}$. For $N=3$ case, we have
\begin{align}
	&\tilde{f}_{2,n;m}(\chi, \sigma;\bar{\chi},\bar{\sigma})\equiv \delta_{\eta_n} \Phi^{\text{\tiny dressed}}_{m} \cr
	=&\gamma_2\;  G_m(z_1,z_2;\bar{z}_1,\bar{z}_2) e^{-{2\pi  i n \chi \over \tau} } \left[n\cos { 2\pi n \sigma\over \tau } -{\sin { 2\pi n\sigma\over \tau}\over \tan {2\pi (\sigma + \pi m )\over \tau }}\right]\\
	&\tilde{f}_{3,n;m}(\chi, \sigma;\bar{\chi},\bar{\sigma})\equiv \delta_{\zeta_n} \Phi^{\text{\tiny dressed}}_{m} \cr
	=&\gamma_3\; G_m(z_1,z_2;\bar{z}_1,\bar{z}_2) e^{-{2\pi  i n \chi \over \tau} } \left[ 2 n^2 \sin { 2\pi n \sigma \over \tau } + 6 n  { \cos { 2\pi n \sigma \over \tau }\over \tan { 2\pi (\sigma+ \pi m) \over \tau }} \right.\cr
	&\hspace{60mm}\left.- 2{ 1+2\cos^2 { 2\pi  (\sigma+ \pi m) \over \tau }\over\sin^2{ 2\pi (\sigma+ \pi m) \over \tau }} \sin { 2\pi n \sigma \over \tau }\right]	
\end{align}
where $\gamma_2$ and $\gamma_3$ are given in \eqref{eq: gamma coefficient}. It is useful to choose a particular configuration $(z_1,\bar{z}_1;z_2,\bar{z}_2)=(\chi-{\tau\over 4},\bar{\chi}-{\bar{\tau}\over 4};\chi+{\tau\over 4},\bar{\chi}+{\bar{\tau}\over 4} )$ or equivalently $(\chi,\sigma,\bar{\chi},\bar{\sigma})=(\chi,-{\tau\over 4},\bar{\chi},-{\bar{\tau}\over 4})$ in order to simplify the OTOC calculation~\cite{Maldacena:2016hyu}. With this choice of configuration, the leading one-point function of the bi-local becomes
\begin{equation}
	G_m(\sigma=-{\tau \over 4},\bar{\sigma}=-{ \bar{\tau} \over 4} )={1\over \left[\cos {2 \pi^2 m\over \tau} \cos {2 \pi^2 m\over \bar{\tau} }\right]^{2h}}\label{eq: two point function image}
\end{equation}
Note that $G_m$ is independent of $\chi$ and $\bar{\chi}$, and we will parametrize it by $\sigma$ and $\bar{\sigma}$. \ie $G_m(\sigma,\bar{\sigma})$. The soft mode eigenfunction $\tilde{f}_{s,n,m}=$ ($s=2,3$) can also be simplified as follow.
\begin{align}
	{\tilde{f}_{2,n,m}(\chi , -{\tau \over 4}; \bar{\chi} ,-{\bar{\tau} \over 4})\over G_{m}(-\tau/ 4,-\bar{\tau}/4) }=&\gamma_2\; e^{- {2\pi i  n \chi\over \tau} }  \left[n\cos {  n \pi \over 2 } -\sin {  n \pi \over 2} \tan { 2\pi^2 m \over \tau }\right]\label{eq: soft mode specific pt2}\\
	{\tilde{f}_{3,n,m}(\chi , -{\tau \over 4}; \bar{\chi} ,-{\bar{\tau} \over 4})\over G_{m}(-\tau/ 4,-\bar{\tau}/4) }=&\gamma_3\; e^{- {2\pi i  n \chi\over \tau} }  \left[-2\left(n^2+2-{3\over \cos^2{2m\pi^2\over \tau}}\right)\sin {  n \pi \over 2 }\right.\cr 
	&\hspace{40mm} \left.-6n\cos {  n \pi \over 2} \tan { 2\pi^2 m \over \tau }\right]\label{eq: soft mode specific pt3}
\end{align}
Note that the last term in \eqref{eq: soft mode specific pt2} and \eqref{eq: soft mode specific pt3} is odd in $m$. Hence, by summing them over $m$, those linear terms will be cancelled. But, the $m$ dependence of the first term in~\eqref{eq: soft mode specific pt3} still survives unlike $s=2$ case. Hence, the soft mode expansion of the dressed bi-locals for $s=2$ mode at $\sigma=-{\tau\over 4}$ can be factorized into a simple form
\begin{align}
    \sum_m	\delta_{\eta_n} \Phi^{\text{\tiny dressed}}_{m}\left(\chi-{\tau\over 4},\bar{\chi}-{\bar{\tau}\over 4};\chi+{\tau\over 4},\bar{\chi}+{\bar{\tau}\over 4} \right)=& \gamma_2\; e^{-{2\pi  i n \chi \over \tau} } n \cos {  n \pi \over 2 }G_{\text{\tiny BTZ}}\left(-\tau/ 4,-\bar{\tau}/4\right)\label{eq: deviation of dressed operator}
\end{align}
while for $s=3$ case we have 
\begin{align}
    &\sum_m	\delta_{\zeta_n} \Phi^{\text{\tiny dressed}}_{m}\left(\chi-{\tau\over 4},\bar{\chi}-{\bar{\tau}\over 4};\chi+{\tau\over 4},\bar{\chi}+{\bar{\tau}\over 4} \right)\cr
    =&-2\gamma_3\; e^{-{2\pi  i n \chi \over \tau} }\sum_m \left(n^2+2-3 \sec^2{2m\pi^2\over \tau}\right)\sin {  n \pi \over 2 }G_{m}\left(-\tau/ 4,-\bar{\tau}/4 \right)\label{eq: deviation of dressed operator1}
\end{align}

We will evaluate the following OTOC which is regularized by thermal density matrix $\rho=e^{-\beta H}$~\cite{Maldacena:2015waa,Maldacena:2016hyu}:
\begin{equation}
    F(t,\phi)= \mathfrak{tr} \left[\rho^{1\over 4} V(0)\rho^{1\over 4} W(t,\phi)\rho^{1\over 4} V(0)\rho^{1\over 4} W(t,\phi) \right]
\end{equation}
For this, we first evaluate the Euclidean correlator in~\eqref{eq: euclidean 4pi for otoc} at the following configuration~\cite{Maldacena:2016hyu}:
\begin{align}
	(z_1,\bar{z}_1)=&(\chi-{\tau\over 4},\bar{\chi}- {\bar{\tau}\over 4} )\\
	(z_2,\bar{z}_2)=&(\chi+{\tau\over 4},\bar{\chi}+{\bar{\tau}\over 4} )\\
	(z_3,\bar{z}_3)=&(0, 0 )\\
	(z_4,\bar{z}_4)=&({\tau\over 2}, {\bar{\tau}\over 2} )
\end{align}
We evaluate the contribution of each spin-$s$ soft mode to the Euclidean four point function in~\eqref{eq: euclidean 4pi for otoc}:
\begin{equation}
    \mathcal{F}_{s}(1,2,3,4)\equiv \left.{F(1,2,3,4)\over G_{\text{\tiny BTZ}}(1,2)G_{\text{\tiny BTZ}}(3,4) }\right|_{\text{\tiny spin-$s$}}=\sum_n \langle \zeta^{(s)}_n\zeta^{(s)}_{-n}  \rangle { \delta_{\zeta_{n}^{(s)}}G(1,2) \delta_{\zeta_{-n}^{(s)}}G(3,4)\over G_{\text{\tiny BTZ}}(1,2)G_{\text{\tiny BTZ}}(3,4) }
\end{equation}
Here, we will use the two point function of soft modes from the on-shell action (\eg \eqref{eq: soft mode 2pt s2} and \eqref{eq: soft mode 2pt s3}).

\begin{figure}[ht!]
\centering
\begin{minipage}[c]{0.4\linewidth}
\centering
    \includegraphics[width=\textwidth]{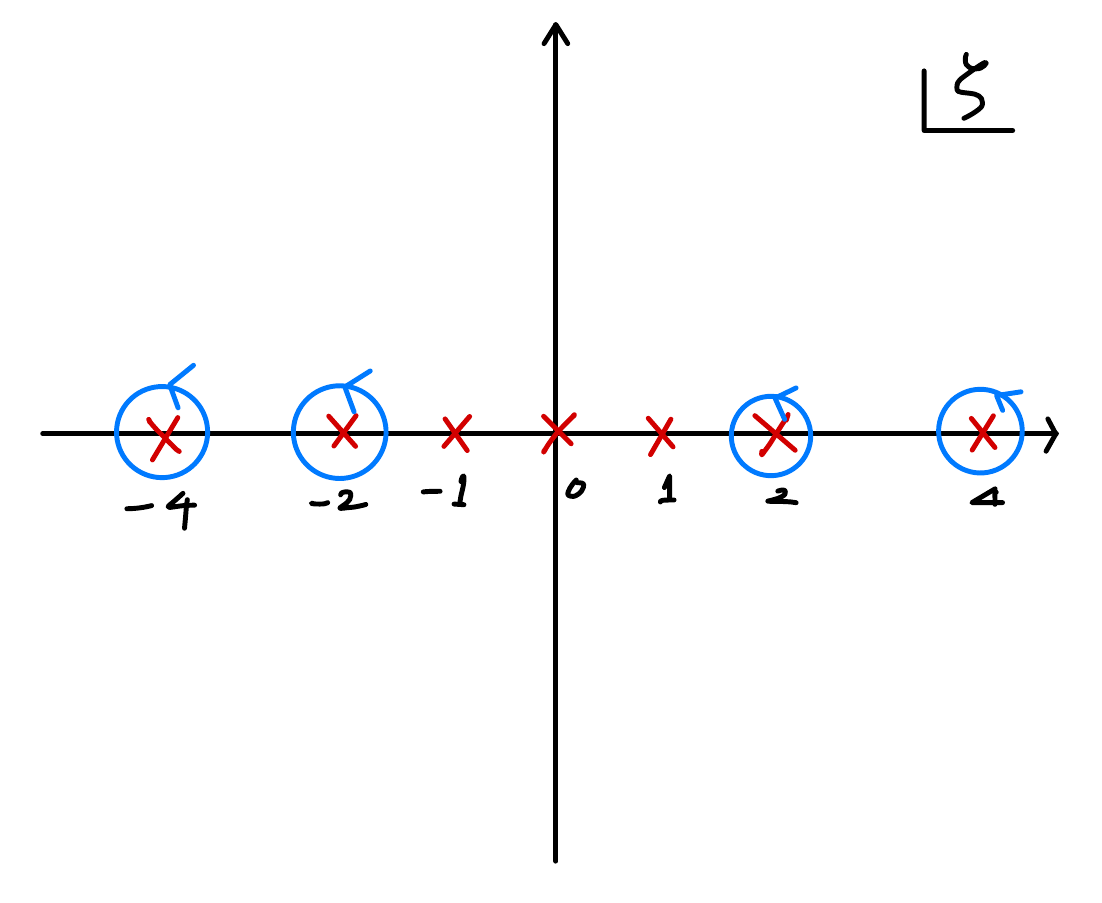}
\end{minipage}
\quad $\Longrightarrow$\quad 
\begin{minipage}[c]{0.4\linewidth}
\centering
    \includegraphics[width=\textwidth]{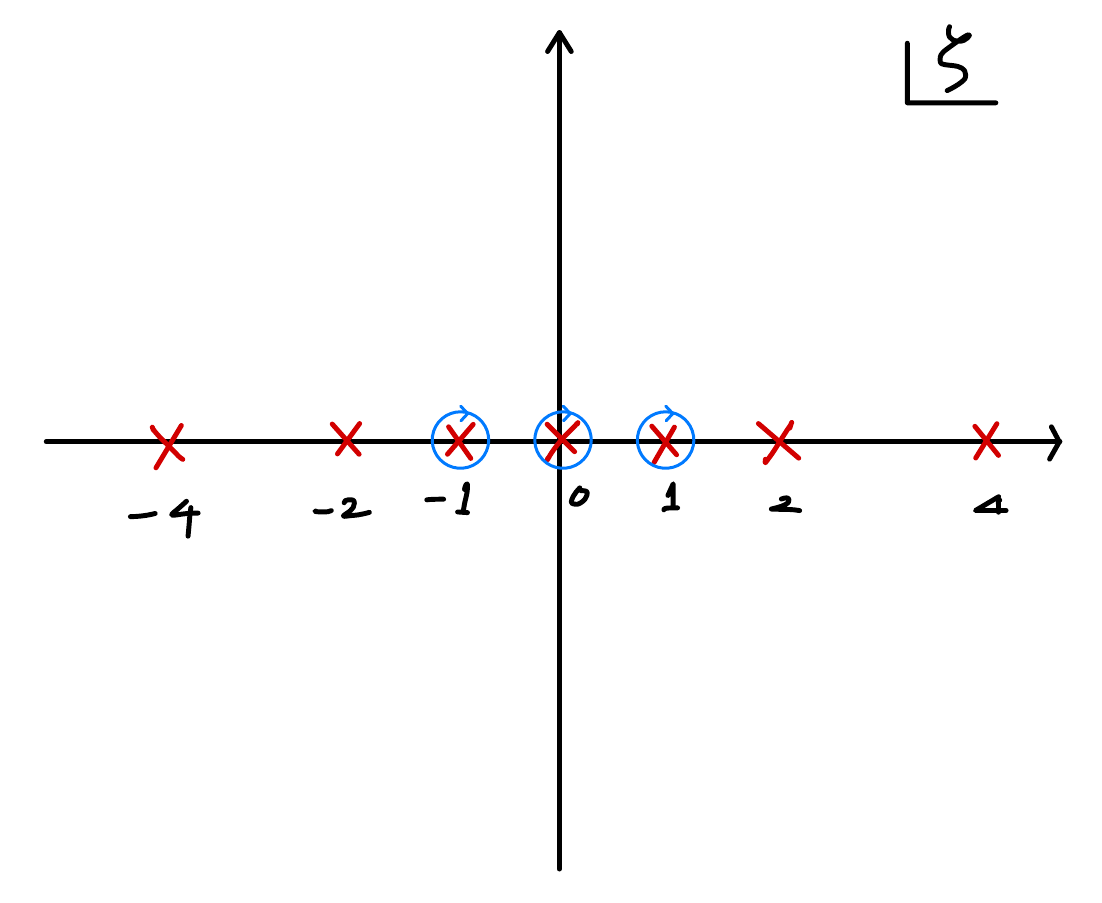}
\end{minipage}
\caption{The contour integral for the graviton contribution.}
	\label{fig1}
\end{figure}

For the graviton contribution ($s=2$), we have
\begin{align}
    \mathcal{F}_2(1,2,3,4)=&-\sum_{|n|= 2,4, 6,\cdots}\gamma_2^2\kappa_2 {(-1)^{n\over 2} \over n^2-1 }e^{-{2\pi  i n \chi \over \tau} } +\text{(anti-holomorphic)}\cr
    =&-\gamma_2^2\kappa_2  {1\over 2\pi i} \oint_{\mathcal{C}_2 } d\zeta {{\pi\over 2} \over \sin {\pi \zeta\over 2} }{ e^{-{2\pi  i \zeta \chi \over \tau}  }\over \zeta^2-1 }+\text{(anti-holomorphic)}
\end{align}
where the summation runs over the even integer due to the special choice of the coordinates. In the second line, we rewrite the infinite series into a contour integral\footnote{See \cite{Sarosi:2017ykf} for clear explanation on the calculation in~\cite{Maldacena:2016hyu,Yoon:2017nig,Narayan:2017hvh} by using contour integral.} where the contour $\mathcal{C}_2$ is a collection of small counterclockwise circle around $\zeta=\pm 2,\pm4,\cdots $. Note that the integrand also has a simple pole at $\zeta=0,\pm1$. Hence, by pushing the contour to infinity, the contour integral becomes the summation of the residue at $\zeta=-1,0,1$ (See Figure~\ref{fig1}):
\begin{align}
    \mathcal{F}_2(1,2,3,4)=&\gamma_2^2\kappa_2   \left[{\pi \over 4} e^{-{2\pi  i \chi \over \tau}  } - 1 +{\pi  \over 4}  e^{{2\pi  i  \chi \over \tau}} \right]+\text{(anti-holomorphic)}
\end{align}
With the Euclidean correlator, we will take the analytic continuation of the Euclidean time $t_E$ into the Lorentzian time $t$
\begin{equation}
    \chi=\phi + i{t_E\over l}= \phi - {t\over l}\ ,
\end{equation}
For simplicity, we will consider non-rotating BTZ black hole\footnote{See~\cite{Jahnke:2019} for rotating BTZ black hole in $SL(2)$ Chern-Simons gravity.} (embedded in higher spin gravity) where the modular parameter is given by\footnote{The OTOCs of the rotating BTZ black hole in the Einstein gravity has been studied where one need ``Boyer-Lindquist co-rotating frame'' for the analytic continuation to real time~\cite{Jahnke:2019}.}
\begin{equation}
    \tau={i\beta\over l}={2\pi i l \over r_+}
\end{equation}
after the analytic continuation, we have
\begin{align}
    \mathcal{F}_2(t,\phi)=& -{48\beta\over \pi^2 l c}   \left[{\pi \over 4} e^{{2\pi   \over \beta}(t-l\phi)  } - 1 +{\pi  \over 4}  e^{-{2\pi   \over \beta}(t-l\phi)} \right]+\text{(anti-holomorphic)}
\end{align}
where we used $c=6\kc$.

In contrast to the graviton contribution, the spin-$3$ field contribution has $m$ dependence even at the special coordinates that we focus on.
\begin{align}
    &\mathcal{F}_{3}(1,2,3,4)\cr
    \equiv&-\sum_{m,m'}\sum_{|n|= 3,5, 7,\cdots}4i\gamma_3^2\kappa_3 {(-1)^{n-1\over 2}\left(n^2+2-3 \sec^2{2m\pi^2\over \tau}\right)\mathcal{A}_m\left(n^2+2-3 \sec^2{2m'\pi^2\over \tau}\right)\mathcal{A}_{m'} \over n^2(n^2-1)(n^2-4) }e^{-{2\pi  i n \chi \over \tau} }  \cr
    &+\text{(anti-holomorphic)}
\end{align}
where we define
\begin{equation}
    \mathcal{A}_m \equiv  {G_{m}(-{\tau\over 4},-{\bar{\tau}\over 4}) \over G_{\text{\tiny BTZ}}(-{\tau\over 4},-{\bar{\tau}\over 4})}
\end{equation}
For $m= 0$, it is analogous to $s=2$ case where we can rewrite the infinite series as contour integral: 
\begin{align}
    \mathcal{F}_{3,m=0}(1,2,3,4)=&-4i\gamma_3^2\kappa_3  {1\over 2\pi i} \oint_{\mathcal{C}_3 } d\zeta {{\pi\over 2} \over \cos {\pi \zeta\over 2} }{\zeta^2-1 \over \zeta^2(\zeta^2-4) }e^{-{2\pi  i \zeta \chi \over \tau}  }+\text{(anti-holomorphic)}
\end{align}
where the contour $\mathcal{C}_3$ is a collection of small counterclockwise circle at $\zeta=\pm3, \pm 5,\cdots$. In the same way, we move the contour to pick up poles outside of the contour $\mathcal{C}_3$. Noting that the integrand has a simple pole at $\zeta=\pm2 $ and double pole at $\zeta=0$, we have
\begin{align}
    \mathcal{F}_{3,m=0}(1,2,3,4)=&4i\gamma_3^2\kappa_3 \left[-{3\pi\over 32} e^{-{4\pi  i \chi \over \tau}  } -{i \pi^2\over 4\tau }\chi +{3\pi\over 32} e^{{4\pi  i \chi \over \tau}  } \right]+\text{(anti-holomorphic)}
\end{align}

For $m\ne 0$, one can write the infinite series into a similar contour integral:
\begin{align}
    &\mathcal{F}_{3,m\ne 0}(1,2,3,4)\cr
    =&-4i\gamma_3^2\kappa_3 \sum_{m,m'} {1\over 2\pi i} \oint_{\mathcal{C}_3 } d\zeta {{\pi\over 2} \over \cos {\pi \zeta\over 2} }{ \left(\zeta^2-1-3 \tan^2{2m\pi^2\over \tau}\right)\mathcal{A}_m \left(\zeta^2-1-3 \tan^2{2m'\pi^2\over \tau}\right) \mathcal{A}_{m'}\over \zeta^2(\zeta^2-1)(\zeta^2-4) }e^{-{2\pi  i \zeta \chi \over \tau}  }\cr
    &+\text{(anti-holomorphic)}
\end{align}
Because of the $m$ dependence, the structure of poles could be changed for $m\ne 0$. We are interested, for simplicity, in (non-rotating) BTZ black hole where the modular parameter is pure imaginary. \ie $\tan({2m \pi^2 \over \tau}) = i \tanh({2m \pi^2 l \over \beta})$.  Then, we have
\begin{equation}
    n^2-1 \leqq n^2-1+3\tanh^2{2m\pi^2 l\over \beta }< n^2+2
\end{equation}
This implies that the $m$ dependence does not change the simple pole at $\zeta=2$ which will lead to the leading exponential growth after analytic continuation.\footnote{For some special $m$, the $\zeta=0$ pole could be changed.} Also, by moving the contour, the spin-$s$ contribution for all $m$ can be written as
\begin{align}
    \mathcal{F}_3(1,2,3,4)=&4i\gamma_3^2\kappa_3 \left[-{3\pi\over 8} \left(e^{-{4\pi  i \chi \over \tau}  }-e^{{4\pi  i \chi \over \tau}  }\right) \mathcal{A}^2 +\cdots  \right]\cr
    &+\text{(anti-holomorphic)}\label{eq: spin 3 nonzero m}
\end{align}
where $\mathcal{A}$ is defined by
\begin{equation}
    \mathcal{A}\equiv \sum_m  p_m \mathcal{A}_m\hspace{5mm}\mbox{where}\quad p_m\equiv {1\over 2}\left(1-\tan^2{2m\pi^2\over \tau}\right)
\end{equation}
Here, we present only the contribution from the pole at $\zeta=\pm 2$. It is also easy to evaluate the contribution from other poles, but they will give the sub-leading exponential growth.

For (non-rotating) BTZ black hole, the modular parameter $\tau$ is pure imaginary\footnote{The AdS vacuum has real modular parameter. \ie $\tau=2\pi$. But, in this case, the angular coordinate is already periodic, and we don't have to sum up those images.} so that we have
\begin{equation}
    0\leqq p_m\leqq 1
\end{equation}
and as $m$ becomes larger, $p_m$ approaches to 1. Note that $\mathcal{A}$ can be viewed as averaged $p_m$ weighted by $G_m$. Using \eqref{eq: two point function image}, one can easily estimate that $\mathcal{A}$ approaches to 1 for $h<0$ which is the conformal dimension of our bi-locals. In the contributions of other poles, which include $e^{\pm {2\pi i \chi \over \tau}}$ $\chi e^{\pm {2\pi i \chi \over \tau}}$ $\chi$, also have a similar $m$ dependence. \eg
\begin{equation}
    \sum_m \tan^2{2m\pi^2\over \tau} \mathcal{A}_m\quad,\quad \sum_m \sec^2{2m\pi^2\over \tau} \mathcal{A}_m\quad,\quad \sum_m \sec^2{2m\pi^2\over \tau}\tan^2{2m\pi^2\over \tau} \mathcal{A}_m
\end{equation}
And, they would also become of order 1 or vanish. After analytic continuation, the leading exponential growth in real time in the spin-$3$ contribution is 
\begin{align}
    \mathcal{F}_3(1,2,3,4)=& {432 i \beta\over \pi c l} \left[ \mathcal{A} e^{{4\pi   \over \beta}(t-l\phi)  }  +\cdots \right]+\text{(anti-holomorphic)}
\end{align}
And, the leading Lyapunov exponent is ${4\pi \over \beta}$.

\section{Conclusion}
\label{sec:conclusion}

In this paper, we investigated the quantum chaos in $SL(N)$ Chern-Simons higher spin gravity. With the boundary term and the corresponding boundary condition in~\cite{deBoer:2013gz}, we have derived the quadratic on-shell action of the smooth higher spin modes in $SL(N,\mathbb{C})$ Chern-Simons higher spin gauge theories in EAdS$_3$. We considered the gravitational dressing of bi-local objects such as the matter master field in $SL(N)$ Vasiliev equation and Wilson line in the non-constant background. From the soft higher spin mode expansion of the dressed bi-locals, we have obtained the soft mode eigenfunction. In addition, based on the null relations in $\mathcal{W}_N$ minimal model, we have shown that the $\mathcal{W}$-Ward identity leads to the same eigenfunction. We have also constructed a recursion relation of the soft mode eigenfunctions, and we found the form of the eigenfunction for all spin $s$. Using the on-shell action and the soft mode eigenfunction, we have evaluated the OTOCs and have explicitly shown that the Lyapunov exponent of the $SL(3)$ Chern-Simons higher spin gravity is ${4\pi \over \beta}$. In Appendix~\ref{app: lyapunov exponent}, we also evaluate OTOCs for the $SL(N)$ case and the Lyapunov exponent is found to be ${2\pi \over \beta}(N-1)$.

For $N>2$, the $SL(N)$ Chern-Simons higher spin gravity violates the chaos bound, which reflect that it is not unitary. The $SL(N)$ higher spin gravity can be considered as the semi-classical limit of $hs[\lambda]$ higher spin gravity which is unitary for $\lambda\in [0,1]$. The $hs[\lambda$ higher spin gravity has infinite tower of higher spin fields, and it was shown in~\cite{Perlmutter:2016pkf} that its Lyapunov exponent is zero. In the semi-classical limit ($\lambda\rightarrow -N$), the infinite tower of higher spin fields is truncated to a finite number of higher spin fields, and at the same time, it becomes non-unitary. We have observed the non-unitarity in the negative conformal dimension of the scalar field and the Wilson line discussed in Section~\ref{sec: zero mode eigenfunction}.

It is also interesting to consider a possibility of the higher spin generalization of the chaos bound. The 2D and 3D gravity can have a finite number of interacting higher spin fields at the cost of losing unitarity and causality~\cite{Perlmutter:2016pkf}. Therefore, in the proof of the bound on chaos, the argument based on the unitarity and the causality will not hold, which leads to the violation of the chaos bound. In spite of the violation of the bound, the presence of a finite number of higher spin fields seemingly suggest a larger bound on Lyapunov exponent. It would be interesting to prove a new bound on chaos in the non-unitary field theories\footnote{Recently, it was reported that the non-unitary Fishnet model violates the bound on chaos~\cite{deMelloKoch:2019ywq}.} of which bulk dual has a finite number of higher spin fields.

In this work, we have considered the simplest constant background without spin-$3$ chemical potential. Therefore, the on-shell action has only the spin-$2$ charge $\mathcal{L}$. If we turn on the spin-$3$ chemical potential, it will generate the spin-$3$ charge $\mathcal{W}$ in the on-shell action. Then, the quadratic action contains the interaction between the spin-$2$ and spin-$3$ soft modes:
\begin{equation}
    \int {dz^2 \over 2\pi \im (\tau) }\mathcal{W}\sim \sum_{n\in \mathbb{Z}} n^2(n^2-1)(n^2-4) f_{-n}g_n 
\end{equation}
This is analogous to the coupling between $U(1)$ and reparametrization modes in complex SYK model and 2D gravity~\cite{Davison:2016ngz}. It was reported in~\cite{David:2017eno} that the presence of the spin-$3$ chemical potential can change entanglement entropy negative, which leads to the bound on spin-$3$ chemical potential. It will be interesting to investigate the quantum chaos with the coupling between spin-$2$ and spin-$3$ soft mode.

In \cite{Haehl:2018izb}, it was pointed out\footnote{We thank Felix Haehl for bringing this to our attention.} that the ``pole-skipping'' phenomenon~\cite{Grozdanov:2017ajz,Blake:2017ris,Haehl:2018izb,Blake:2018leo,Grozdanov:2018kkt} of the $W_3$ higher spin current can capture the Lyapunov exponent ${4\pi \over \beta}$. These techniques might provide alternate ways to understanding the Lyapnov exponent in the context of higher spin holography.

\acknowledgments

We would like to thank Antal Jevicki, Matthias R. Gaberdiel, Mikhail A. Vasiliev, Spenta Wadia, Robert de Mello Koch, Euihun Joung, Thomas Basile, Keun-Young Kim, Viktor Jahnke, Sangjin Sin, Dario Rosa and especially Ioannis Papadimitriou for extensive discussions. JY thanks the Korea Institute for Advanced Study (KIAS) for the hospitality and support during the initial stages of this work, within the program ``IPMU-KIAS-Kyunghee Univ. joint workshop 2017''. JY thank the Yukawa Institute for Theoretical Physics at Kyoto University for generous support during the course of this work, within the workshop YITP-T-18-04 ``New Frontiers in String Theory 2018''. JY thank the Kavli Institute for Theoretical Physics for generous support during the course of this work, within the workshop ``Chaos and Order: from strongly correlated systems to black holes 2018''. This research was supported in part by the National Science Foundation under Grant No. NSF PHY-1748958. JY thank the Okinawa Institute of Science and Technology~(OIST) for the hospitality and generous support during the course of this work, within the workshop ``OIST Mini Symposium, Holographic Tensors 2018''. JY thanks the Erwin Schrodinger International Institute (ESI) where this work was completed during the program ``Higher Spins and Holography 2019'', and JY would like to thank the organizers for giving an opportunity to present our work.


\appendix

\section{Conventions}
\label{app: convention}

In this paper, we use the following representation for the generators of $sl(3)$:
\begin{equation}\label{eq:sl3 L generators}
	L_1=\begin{pmatrix}
	0 & 0 & 0\\
	\sqrt{2} & 0 & 0\\
	0 & \sqrt{2} & 0\\
	\end{pmatrix}\hspace{5mm},\hspace{5mm} L_0=\begin{pmatrix}
	1 & 0 & 0\\
	0 & 0 & 0\\
	0 & 0 & -1\\
	\end{pmatrix}\hspace{5mm},\hspace{5mm} L_{-1}=\begin{pmatrix}
	0 & -\sqrt{2} & 0\\
	0 & 0 & -\sqrt{2}\\
	0 & 0 & 0\\
	\end{pmatrix}
\end{equation}
\begin{align}
	W_2=\begin{pmatrix}
	0 & 0 & 0\\
	0 & 0 & 0\\
	4 & 0 & 0\\
	\end{pmatrix}\hspace{5mm},\hspace{5mm} W_1=\begin{pmatrix}
	0 & 0 & 0\\
	\sqrt{2} & 0 & 0\\
	0 & -\sqrt{2} & 0\\
	\end{pmatrix}\hspace{5mm},\hspace{5mm} W_0={2\over 3} \begin{pmatrix}
	1 & 0 & 0\\
	0 & -2 & 0\\
	0 & 0 & 1\\
	\end{pmatrix}
\end{align}
\begin{align}\label{eq:sl3 W generators}
	W_{-1}=\begin{pmatrix}
	0 & -\sqrt{2} & 0\\
	0 & 0 & \sqrt{2}\\
	0 & 0 & 0\\
	\end{pmatrix}\hspace{5mm},\hspace{5mm} W_{-2}=\begin{pmatrix}
	0 & 0 & 4\\
	0 & 0 & 0\\
	0 & 0 & 0\\
	\end{pmatrix}
\end{align}
and, their traces are 
\begin{equation}
	\tr (L_n L_{-m})= \begin{pmatrix}
	0 & 0 & -4\\
	0 & 2 & 0\\
	-4 & 0 & 0\\
	\end{pmatrix}\hspace{5mm},\hspace{5mm} \tr(W_nW_{-m}) =\begin{pmatrix}
	0 & 0 & 0 & 0 & 16\\
	0 & 0 & 0 & -4 & 0 \\
	0 & 0 & {8\over 3} & 0 & 0\\
	0 & -4 & 0 & 0 & 0\\
	16 & 0 & 0 & 0 & 0\\
	\end{pmatrix}
\end{equation}
Note that this trace $\tr$ is the summation over the fundamental representation. The normalized trace $\Tr$ is defined by
\begin{equation}
    \Tr(A)={1\over \tr(L_0L_0)} \tr(A)
\end{equation}

For $SL(2,\mathbb{C})$ Chern-Simons gravity, the representations of the generators of $sl(2)$ are
\begin{equation}
	L_1=\begin{pmatrix}
	0 & 0 \\
	1 & 0 \\
	\end{pmatrix}\hspace{5mm},\hspace{5mm} L_0=\begin{pmatrix}
	{1\over 2} & 0 \\
	0 & -{1\over 2} \\
	\end{pmatrix}\hspace{5mm},\hspace{5mm} L_{-1}=\begin{pmatrix}
	0 & -1\\
	0 & 0 \\
	\end{pmatrix}\hspace{5mm},\hspace{5mm} 
\end{equation}

\section{Conjecture on Null Relation and $\mathcal{W}$ transformation }
\label{app: infinitesimal transf}

In this appendix, we make a conjecture generalizing the null relation discussed in Section~\ref{sec: ward identity}. Based on this conjecture, we derive the infinitesimal transformation of two point function under the $W_s$ transformation.

In large $c$ limit of $\mathcal{W}_N$ minimal model, we conjecture that for a primary operator $\phi$, the action of $W^{(s)}_{-n}$ on $\phi$ will be proportional to $(L_{-1})^n\phi$ in large $c$ limit:
\begin{equation}
    W^{(s)}_{-n} \phi\sim (L_{-1})^n\phi +\mathcal{O} (1/c) 
\end{equation}
Note that in this conjecture we consider special primary operator in $\mathcal{W}_N$ minimal model such as $({\tiny \yng(1)};0)$, $({\tiny \yng(1,1)};0)$ \etc. But, this conjecture might hold in a large class of primary operators in $\mathcal{W}_N$ minimal model (\eg simple primaries constructed on the Fock space found in~\cite{Chang:2011vka,Jevicki:2013kma,Chang:2013izp}). Motivated by the explicit expressions for the low $s$,  we consider the following ansatz.
\begin{align}
	W^{(s)}_{-n} \phi=& {a_n w^{(s)}\over \prod_{j=0}^{n-1} (2h+j) } (L_{-1})^n\phi 
	\label{eq: w action on primary}
\end{align}
where $h$ and $w^{(s)}$ denotes the $L_0$ and $W^{(s)}_0$ charge of the primary $\phi$, respectively. \ie we will choose $a_0=1$.  
%
%
Here, we chose  To find $a_n$, we act $L_1$ on the both sides of \eqref{eq: w action on primary}. Using the commutation relation
\begin{equation}
	[L_m,W^{(s)}_n]=( (s-1)m-n)W_{m+n}\ ,
\end{equation}
we have
\begin{equation}
	L_1 W^{(s)}_{-n} \phi=(s-1+n)W_{-n+1}^{(s)}  \phi= {a_n w^{(s)}\over \prod_{j=0}^{n-1} (2h+j) } L_1 L_{-1}^n\phi={n a_n w^{(s)}\over \prod_{j=0}^{n-2} (2h+j) } L_{-1}^{n-1} \phi
\end{equation}
This leads to recursion relation of $a_n$, and we have
\begin{equation}
	a_n={s-1+n\over n}a_{n-1}=\cdots = {(s-1+n)! \over n! (s-1)!}a_{0}={(s-1+n)! \over n! (s-1)!}
\end{equation}
Hence, based on our conjecture, the action of $W^{(s)}_{-n}$ on the primary operator $\phi$ in large $c$ is found to be 
\begin{align}
	W^{(s)}_{-n} \phi= {(s-1+n)! \over n! (s-1)!} { (2h+n) w^{(s)}\over \prod_{j=0}^{n} (2h+j) } (L_{-1})^n\phi
\end{align}
Note that for the given $\mathcal{W}_s$ charge of $\phi$, the $\mathcal{W}_s$  charge of the conjugate $\overline{\phi}$ is given by
\begin{align}
	W^{(s)}_0\phi=w^{(s)}\phi\quad\Longrightarrow \quad W^{(s)}_0\overline{\phi}=(-1)^s w^{(s)} \overline{\phi}
\end{align}
Using the action of $W^{(s)}_{-n}$ generators on the primary, one can find the $\mathcal{W}_s$ transformation of two point function at zero temperature as in Section~\ref{sec: ward identity}:
\begin{align}
	&{\delta_{\epsilon} \langle \phi(z_1)\overline{\phi}(z_2) \rangle\over \langle \phi(z_1)\overline{\phi}(z_2) \rangle }\cr
	=&-\sum_{m=0}^{s-1} {1\over (s-1-m)!} \left(  \partial_1^{s-1-m}\epsilon(z_1)  \langle (W^{(s)}_{-m}\phi)(z_1)\phi(z_2)\rangle + \partial_2^{s-1-m}\epsilon(z_2) \langle \phi(z_1)(W^{(s)}_{-m}\phi)(z_2)\rangle\right)\cr
	=&- {w^{(s)}\over (s-1)!}\sum_{n} \epsilon_n \sum_{m=0}^{s-1}  { (n+s-1)!(s+m-1)!  \over  (n+m)! m!(s-m-1)! }    (-1)^m { z_1^{m+n}  +(-1)^{m+s} z_2^{m+n}  \over (z_1-z_2)^{m}}
\end{align}
where we expanded $\epsilon(z)$ as 
\begin{equation}
	\epsilon(z)=\sum_n z^{n+s-1} \epsilon_n\ ,
\end{equation}
Using the transformation into finite temperature\footnote{In the appendix, we also use $\tau=2\pi $ unit for simplicity, and, we will recover it if necessary.}
\begin{equation}
	z\qquad\Longrightarrow \qquad e^{-i z}\ ,
\end{equation}
we have
\begin{align}
	{\delta_{\epsilon} \langle \phi(z_1)\overline{\phi}(z_2) \rangle\over \langle \phi(z_1)\overline{\phi}(z_2) \rangle }=&- {w^{(s)}\over (s-1)!}\sum_{n} \epsilon_n e^{-in{z_1+z_2\over 2}} \sum_{m=0}^{s-1}  { (n+s-1)!(s+m-1)!  \over  (n+m)! m!(s-m-1)! }    (-1)^s \cr
	&\hspace{35mm}\times { \left(e^{i(m+n){z_1-z_2\over 2} }+(-1)^{m+s} e^{-i (m+n) {z_1-z_2\over 2} }  \right)   \over (-2i \sin {z_1-z_2\over 2})^{m} }\label{eq: transformation at finite temperature general s} \\
	=&- {w^{(s)}\over (s-1)!}\sum_{n} \epsilon_n e^{-in\chi} {(n+s-1)!\over n!} \left[(-1)^s e^{in \sigma}{}_2F_1(1-s,s,1+n,{e^{i\sigma}\over e^{i\sigma}-e^{-i\sigma}})  \right.\cr
	&\hspace{20mm} \left.+e^{-in\sigma} {}_2F_1(1-s,s,1+n,-{e^{-i\sigma}\over e^{i\sigma}-e^{-i\sigma}}) \right] \label{eq:Expression via Ward Identities}
\end{align}  
For $s> |n|$, the above expression vanishes because
\begin{align}
	&  (-1)^s e^{inz}{}_2F_1(1-s,s,1+n,{e^{i\sigma}\over e^{i\sigma}-e^{-i\sigma}})  +e^{-in\sigma} {}_2F_1(1-s,s,1+n,-{e^{-i\sigma}\over e^{i\sigma}-e^{-i\sigma}})  \cr
	\propto& \left[(-1)^s(-1)^{n\over 2} P_{s-1}^{-n}(i \cot[\sigma])+(-1)^{-{n\over 2}}P_{s-1}^{-n}(-i \cot[\sigma])\right]\cr
	& = 0
\end{align}
Ofcourse, this is exactly agree with what we expect. Namely, the soft mode eigenfunction $f_{s,n}$ vanishes when it is involved with global subalgebra \ie $|n|<s$. For $s\leqq |n|$, one can rewrite \eqref{eq:Expression via Ward Identities} in terms of Jacobi polynomial:
\begin{align}
	&{\delta_{\epsilon} \langle \phi(\chi+\sigma)\overline{\phi}(\chi-\sigma) \rangle\over \langle \phi(\chi+\sigma)\overline{\phi}(\chi-\sigma) \rangle }\cr
	=&- {w^{(s)}\over (s-1)!}\sum_{n} \epsilon_n e^{-in\chi} (s-1)! (-1)^s\left[e^{in\sigma}P_{s-1}^{n,-n}(i \cot \sigma)-e^{-in \sigma} P_{s-1}^{-n,n}(i \cot \sigma)\right]
\end{align}
To see the relation to the eigenfunction in large $q$ limit (together with $v\rightarrow 1$ limit) found in~\cite{Maldacena:2016hyu}, we rewrite them using the transformation of hypergeometric function:
\begin{align}
	&f_{s,n}(\chi,\sigma)= e^{-in\chi} {(n+s-1)!\over n!} \left[(-1)^s e^{in\sigma}{}_2F_1(1-s,s,1+n,{e^{i\sigma}\over e^{i\sigma}-e^{-i\sigma}})  \right.\cr
	&\hspace{60mm}\left.+e^{-in\sigma} {}_2F_1(1-s,s,1+n,-{e^{-i\sigma}\over e^{i\sigma}-e^{-i\sigma}}) \right]\\
	=&\sqrt{\pi}  e^{-in\chi} (n+s-1)!(-1)^s e^{\pi i n\over 2} {(\sin \sigma)^s\over s^n}\cr
	&\hspace{20mm}\times \left[ {{}_2F_1({s+n\over 2},{s-n\over 2},{1\over 2}, \cos^2 \sigma)\over \Gamma({n+s+1\over 2})\Gamma({n-s+2\over 2})}-2i\cos \sigma {{}_2F_1({s+n+1\over 2},{s-n+1\over 2},{3\over 2}, \cos^2 \sigma) \over \Gamma({n+s\over 2})\Gamma({n-s+1\over 2})}  \right]\cr
	&+\sqrt{\pi}  e^{-in\chi} (n+s-1)! e^{-{\pi i n\over 2}} {(\sin \sigma)^s\over s^n} \cr
	&\hspace{20mm}\times\left[ {{}_2F_1({s+n\over 2},{s-n\over 2},{1\over 2}, \cos^2 \sigma)\over \Gamma({n+s+1\over 2})\Gamma({n-s+2\over 2})}+2i\cos \sigma {{}_2F_1({s+n+1\over 2},{s-n+1\over 2},{3\over 2}, \cos^2 \sigma) \over \Gamma({n+s\over 2})\Gamma({n-s+1\over 2})}  \right]
\end{align}
For even $s$, one can simplify them as
\begin{align}
	&f_{s,n}(\chi,\sigma)
	=2\sqrt{\pi}  e^{-in\chi} (n+s-1)!(-1)^s {(\sin \sigma)^s\over s^n} \cr
	&\times \left[ \cos {\pi i n\over 2} {{}_2F_1({s+n\over 2},{s-n\over 2},{1\over 2}, \cos^2 \sigma)\over \Gamma({n+s+1\over 2})\Gamma({n-s+2\over 2})}+2\sin {\pi i n\over 2}\cos z {{}_2F_1({s+n+1\over 2},{s-n+1\over 2},{3\over 2}, \cos^2 \sigma) \over \Gamma({n+s\over 2})\Gamma({n-s+1\over 2})}  \right]
\end{align}
For odd$s$, we have
\begin{align}
	&f_{s,n}(\chi,\sigma)
	=2\sqrt{\pi}  e^{-in\chi} (n+s-1)!(-1)^s {(\sin \sigma)^s\over s^n} \cr
	&\times \left[ \sin {\pi i n\over 2} {{}_2F_1({s+n\over 2},{s-n\over 2},{1\over 2}, \cos^2 \sigma)\over \Gamma({n+s+1\over 2})\Gamma({n-s+2\over 2})}-2\cos {\pi i n\over 2}\cos \sigma {{}_2F_1({s+n+1\over 2},{s-n+1\over 2},{3\over 2}, \cos^2 \sigma) \over \Gamma({n+s\over 2})\Gamma({n-s+1\over 2})}  \right]
\end{align}

\section{Recursion relation, Orthogonality and Normalization}
\label{app: recursion}

In this appendix, we will show that the combination of Jacobi polynomials given in~\eqref{eq:g modes not normalized} satisfies all the conditions that we demand. \ie (\textit{i}) They satisfy the recursion relation in~\eqref{eq: recursion relation} (which, of course, means the condition in~\eqref{eq:Zero Mode Condition 1} is satisfied) (\textit{ii}) They are orthogonal to each other~\eqref{eq:Zero Mode Condition 2} with respect to the inner product defined in~\eqref{eq:inner product on g}. We will also determine the normalization of the modes. 

\vspace{3mm}

\noindent
{\bf Proof of Recursion Relation}\;\; The Jacobi Functions $P_{n}^{\alpha,\beta}(u)$ for special arguments $\alpha= -\beta$ satisfy the following recursion relations (See eq.~(2.34) and eq.~(2.32) in \cite{W_nsche_2017})
\begin{equation}
\begin{split}
\left( u (s+1) +n - (1-u^2) \partial_u   \right) P_s^{n,-n}(u) &= (s+1) P_{s+1}^{n,-n} \\
P_{s+1}^{n,-n}(u) & = {2s+1 \over s+1 } u P_s^{n,-n}(u) - {(s^2 - n^2) \over s(s+1)} P_{s-1}^{n,-n}(u)
\end{split}
\end{equation}
Going to variables  $u \equiv i \cot \sigma$ and defining $h_{s,n}(\sigma) \equiv {e^{i n \sigma} P_{s-1}^{n,-n}(i \cot \sigma) \over \sin \sigma}$ and using the above two recursion relations we get  
\begin{equation}
- (s+1) h_{s+2}(\sigma) = i \ {(2s+1) \over s+1} \partial_z h_{s+1}(\sigma) + {n^2- s^2 \over s+1} h_{s,n}(\sigma)
\end{equation}
Defining $H_{s,n}(\sigma) \equiv {(s-1)! (s-1)! \over (2s-3)!} e^{i \pi s \over 2} h_{s,n}(\sigma) $, the above equation can be seen to be
\begin{equation}
H_{s+2,n}(\sigma) = \partial_\sigma H_{s+1,n}(\sigma) + {s^2(n^2- s^2) \over 4s^2-1} H_{s,n}(\sigma)
\end{equation}
If the above recursion is true for $H_{s,n}(\sigma)$, it is also true for $H_{s,-n}(\sigma)$. And therefore it also holds for the linear combination $g_{s,n}(\sigma) = H_{s,n}(\sigma) + H_{s,-n}(\sigma)$ and this completes the proof. 

\vspace{3mm}

\noindent
{\bf Orthogonality}\;\; To show that the functions $g_{s,n}(\sigma)$ is orthogonal, notice that it follows from the differential equation satisfied by Jacobi polynmial that the function $G_{s,n}(\sigma) \equiv e^{in\sigma} P_{s-1}^{n,-n}(\sigma) - e^{-in\sigma} P_{s-1}^{-n,n}(\sigma) $ is a solution of a differential equation
\begin{equation}
	\left[n^2+\partial_\sigma^2-{s(s-1)\over \sin^2 \sigma}\right]G_{s,n}(\sigma)=0
\end{equation}
Since $\int {d\sigma \over \sin^2 \sigma} G_{s,,n}^\ast (\sigma) G_{s',n'}(\sigma) \propto \langle  g_{s,n} | g_{s',n}\rangle$, we can plug the above relations into the equality 
\begin{equation}
	\int d\sigma \; G_{s,n}^\ast (\chi,\sigma)\left(n^2+\partial_\sigma^2\right) G_{s',n'}(\chi,\sigma)=\int d\chi d\sigma \; \left(n^2+\partial_\sigma^2\right) G_{s,n}^\ast (t,z)G_{s',n'}(\chi,\sigma)
\end{equation}
to obtain $	s'(s'-1)\langle g_{s,n} | g_{s',n'}\rangle = s(s-1)\langle g_{s,n} |  g_{s',n'}\rangle$. Hence, we have
\begin{equation}
	\langle g_{s,n}  | g_{s',n'}\rangle=0 \hspace{20mm} \forall s\ne s'
\end{equation}

\vspace{3mm}

\noindent
{\bf Normalization}\;\; Finally, we can now determine the normalization of the modes $N_{s,n} \equiv \langle g_{s,n} | g_{s,n} \rangle$ in a purely algebraic way. Noting that 
\begin{equation}\begin{split}
    0 \equiv & \langle g_{s+2,n} | g_{s,n} \rangle =  \langle \partial_\sigma g_{s+1,n} + F_{s,n} g_{s,n}| g_{s,n} \rangle=-\langle g_{s+1,n}|\partial_\sigma g_{s,n}\rangle +F_{s,n} N_{s,n} \cr
	=&-\langle g_{s+1,n}|g_{s+1,n}-F_{s-1,n}g_{s-1,n}\rangle +F_{s,n} N_{s,n} \cr
	=& N_{s,n} F_{s,n} -N_{s+1,n} 
\end{split}\end{equation}
This gives 
\begin{align}
    N_{s,n} =& N_{s-1,n}  F_{s-1,n} = N_{s-2,n}  F_{s-2,n}   F_{s-1,n} = \dots \cr
    =& n\pi (2s-1)\left[{(s-1)!\over (2s-1)!! }\right]^2 \prod_{j=1}^{s-1} (n^2-j^2)
\end{align}
where we used $N_{1,n}=n\pi$.

\section{Lyapunov Exponent from Spin-$s$ Soft Mode}
\label{app: lyapunov exponent}

In this appendix, we will estimate the contribution of spin-$s$ soft mode to the OTOC based on the soft mode eigenfunctions in Appendix~\ref{app: infinitesimal transf}. For simplicity, we do not consider the non-trivial contribution from non-contractible cycle.\footnote{We guess that they would not change the Lyapunov exponent, but change the overall coefficient of exponential growth by order 1.} 

Recall that the essence of the long time limit of OTOCs lies in the following formula:
\begin{align}
	\mathcal{F}_s\sim \sum_{|n|\geqq s}{1\over \prod_{j=0}^{s-1} (n^2-j^2)} f_{s,n} (\chi,\sigma)f_{s,-n} (\chi',\sigma')\label{eq: euclidean otoc in app}
\end{align}
where $f_{s,n}(\chi,\sigma)$ is the soft mode eigenfunction discussed in Appendix~\ref{app: infinitesimal transf}
\begin{equation}
	f_{s,n}  (\chi,\sigma) = e^{-i n\chi} g_{s,n}(z)
\end{equation}
and
\begin{equation}
	g_{s,n}(\sigma)=0 \hspace{5mm} (|n|=0,1,\cdots, s-1)
\end{equation}
In the calculation of OTOCs, it is useful to choose a special configuration. For this, we evaluate the soft mode eigenfunction at $z_1=\chi$ and $z_2=\chi+\pi$, or equivalently\footnote{If we recover $\tau$, it corresponds to $\sigma=-{\tau\over 4}$.} at~$\sigma=-{\pi\over 2}$:
\begin{align}
	&{\delta_{\epsilon} \langle \phi(0)\overline{\phi}(\pi) \rangle\over \langle \phi(0)\overline{\phi}(\pi) \rangle }\cr
	=&- {w^{(s)}\over (s-1)!}\sum_{n} \epsilon_n e^{-in\chi} \sum_{m=0}^{s-1}  { (n+s-1)!(s+m-1)!  \over  (n+m)! m!(s-m-1)! }    (-1)^{s+m} { \left(e^{-i{n\pi \over 2}}+(-1)^{s} e^{i  {n\pi \over 2} }  \right)   \over 2^m  }
\end{align}  
For even $s$, this simply becomes
\begin{align}
	{\delta_{\epsilon} \langle \phi(\tau_1)\overline{\phi}(\tau_2) \rangle\over \langle \phi(\tau_1)\overline{\phi}(\tau_2) \rangle }
	=- {w^{(s)}\over (s-1)!}\sum_{n} \epsilon_n e^{-in{t_1+t_2\over 2}} \cos \left({n\pi \over 2} \right) \prod_{m=1}^{s-1\over 2 } [n^2-(2m-1)^2]\label{eq: even soft mode at point}
\end{align}  
%
%
%
and, for odd $s$, we have
\begin{align}
	{\delta_{\epsilon} \langle \phi(\tau_1)\overline{\phi}(\tau_2) \rangle\over \langle \phi(\tau_1)\overline{\phi}(\tau_2) \rangle }
	=-i {w^{(s)}\over (s-1)!}\sum_{n} \epsilon_n e^{-in{t_1+t_2\over 2}}   \sin \left({n\pi \over 2}\right) n \prod_{m=2}^{s\over 2 } [n^2-(2m-2)^2]\label{eq: odd soft mode at point}
\end{align}

As in Section~\ref{sec: lyapunov exponent}, we will evaluate the Euclidean four point function in~\eqref{eq: euclidean otoc in app} at the following points:
\begin{equation}
	(\chi,\sigma;\chi',\sigma')=(\chi,-{\pi\over 2};-{\pi\over 2},-{\pi\over 2})
\end{equation}
One can easily see from \eqref{eq: even soft mode at point} and \eqref{eq: odd soft mode at point} that $g_{s,n}(- {\pi\over 2})$ vanishes at $n-s\equiv 1$~$(\text{mod}\; 2)$ \ie
\begin{equation}
	g_{s,n}(- {\pi\over 2})=0\qquad \mbox{for }\quad n=\pm(s+1), \pm(s+3) ,\pm(s+5),\cdots
\end{equation}
Furthermore, the other non-vanishing functions become
\begin{equation}
	g_{s,n}(- {\pi\over 2})\sim\begin{cases}
	\quad n (n^2-2^2)(n^2-4^2)\times\cdots\times [n^2-(s-2)^2]& \quad\mbox{for}\quad s\;\;\text{: even}\\
	\quad (n^2-1)(n^2-3^2)\times\cdots\times [n^2-(s-2)^2]& \quad\mbox{for}\quad s\;\;\text{: odd}\\
	\end{cases}
\end{equation}
Here, we omit the numerical coefficients which depend on spin $s$, but they will not have any influence on the Lyapunov exponent in our case. For even $s$, one can express the infinite series as a contour integral
\begin{align}
	\mathcal{F}_s(\chi)\sim
	{1\over 2\pi i } \oint_{\mathcal{C} } {{\pi \over 2}\over \sin {\pi \zeta\over 2}} {(\zeta^2-2^2)\times \cdots\times [\zeta^2-(s-2)^2]\over (\zeta^2-1)\times \cdots \times [\zeta^2-(s-1)^2] } e^{-i\zeta\chi }
\end{align}
where $\mathcal{C}_s$ is a sum of counterclockwise circles centered at $\zeta=\pm s, \pm(s+2),\pm (s+4),\cdots$ with small radius. By deforming the contour, it can be changed into a contour integral along $\mathcal{C}'$ which is a collection of clockwise circle centered at simple poles $\zeta=0,\pm 1,\pm 3,\cdots, \pm (s-1)$. Hence, evaluating the residues, we have
\begin{align}
	\mathcal{F}_s(\chi)
	\sim & \sum_{\substack{m=0\\n=-s+1+2m}}^{s-1} a_{s,n}e^{-in\chi}
\end{align}
where $a_{s,n}$ is a constant of order $\mathcal{O}(1/c)$. For real time OTOC, we take the analytic continuation
\begin{equation}
    \chi={2\pi \over \tau}(\phi+i {t_E\over l}) \quad\Longrightarrow \quad {2\pi \over i\beta} (l \phi - t)
\end{equation}
Here, we retrieve the modular parameter $\tau={i\beta\over l}$ for non-rotating BTZ black hole. Then, the contribution of spin-$s$ field to OTOC is found to be
\begin{align}
	\mathcal{F}_s(t)
	\sim & \sum_{\substack{m=0\\n=-s+1+2m}}^{s-1} a_{s,n}e^{-{2\pi \over \beta}n (l\phi-t) }=a_{s,s-1}e^{{2\pi \over \beta}(s-1) (t-l\phi)}+ a_{s,s-3}e^{{2\pi \over \beta}(s-2) (t-l\phi)}+\cdots 
\end{align}
Then, one can easily read off the leading exponential growth, and the Lyapunov exponent is 
\begin{equation}
	\lambda_s={2\pi(s-1)\over \beta}
\end{equation}
Also, one can repeat the same analysis for odd $s$, and one can obtain a similar result. That is, the OTOCs will be written as follow.
\begin{align}
	\mathcal{F}_s(\chi)\sim
	{1\over 2\pi i } \oint_{\mathcal{C}' } {{\pi \over 2}\over \cos {\pi \zeta\over 2}}{ (\zeta^2-1)\times \cdots \times [\zeta^2-(s-2)^2]\over \zeta^2 (\zeta^2-2^2)\times \cdots\times [\zeta^2-(s-1)^2] }  e^{-i\zeta\tau }
\end{align}
where the deformed contour $\mathcal{C}'$ is a collection of clockwise circle centered at $s=0,\pm 2,\cdots, \pm (s-1)$. Note that in contrast to the previous even $s$, the integrand for odd $s$ has double pole at $\zeta=0$, and this would give extra contribution to the OTOCs. After analytic continuation to real time, we have
\begin{align}
	\mathcal{F}_s\sim& \sum_{\substack{m=0\\n=-s+1+2m}}^{{s-3\over 2}} a_{s,n}e^{-{2\pi \over \beta}n (l\phi-t) } + a_{s,0}{2\pi \over \beta}(t-l\phi) + \sum_{\substack{m={s+1\over 2}\\n=-s+1+2m}}^{s-1} a_{s,n}e^{-{2\pi \over \beta}n (l\phi-t) } \cr
	=&a_{s,s-1}e^{{2\pi \over \beta}(s-1) (t-l\phi)}+ a_{s,s-2}e^{{2\pi \over \beta}(s-2) (t-l\phi)}+\cdots+  a_{s,0}{2\pi \over \beta}(t-l\phi) + \cdots 
\end{align}
Note that the linear growth comes from the double pole at $\zeta=0$. The Lyapunov exponent is as before
\begin{equation}
	\lambda_s= {2\pi (s-1)\over \beta}
\end{equation}
But, for $s=0$, there is no exponential growth, and therefore, the Lyapunov exponent is zero, but the OTOC grows linearly in time~\cite{Yoon:2017nig,Choudhury:2017tax}.

\section{OTOC in $2D$ Higher Spin Gravity}
\label{app: 2d gravity}

In 2D gravity, BF theory with $SL(N,\mathbb{R})$ can describe 2D higher spin gravity~\cite{Vasiliev:1995sv,Alkalaev:2013fsa,Gonzalez:2018enk}, which is analogous to the three-dimensional Chern-Simons higher spin theory. Unlike Chern-Simon gravity, BF theory has only one $SL(N,\mathbb{R})$ gauge connection, but it has extra player, $sl(N)$ Dilaton field $\Phi$. The action of BF theory is given by
\begin{equation}
    S_{\text{\tiny BF}}= \int  \tr [\Phi (dA +A\wedge A)]
\end{equation}
where $A$ and $\Phi$ is gauge connection and dilaton field belonging to $sl(N)$. The equation of motion is given by
\begin{equation}
    dA+ A\wedge A=0\hspace{4mm},\hspace{6mm} d\Phi + [A,\Phi]=0
\end{equation}
One can fix a gauge~\cite{Gonzalez:2018enk} as in the 3D Chern-Simons gravity:
\begin{equation}
    A=b^{-1}(r)(d + a(\tau)d\tau )b(r)
\end{equation}
where we define
\begin{equation}
    b\equiv e^{r L_0}
\end{equation}
Like the asymptotic AdS condition for 3D, one can write the asymptotic AdS condition as
\begin{equation}
    a= L_1 (-1)^s \sum_{s=2}^N \mathcal{W}_{s}(\tau) W^{(s)}_{-s+1}
\end{equation}

In BF theory, we are also interested in a fixed constant background $a_\constb$ and fluctuations around it. \ie
\begin{equation}
    a= h^{-1}(\tau) (a_\constb + \partial_\tau)h(\tau)
\end{equation}
where $h(\tau)$ is a smooth gauge transformation connected to identity. Since the story about the gauge connection is almost the same as that of Chern-Simons gravity, let us focus on the dilaton field. The equation of motion of dilton with constant background $a_\constb$ is given by
\begin{equation}
	\partial_\tau \varphi + [a_\constb, \varphi]=0
\end{equation}
and, its solution is easily found to be
\begin{equation}
	\varphi= e^{-a_\constb t} c e^{a_\constb t} \label{eq: dilaton solution isometry}
\end{equation}
where $c$ is a constant $sl(N)$ matrix. Now, we consider the dilaton with a non-constant background $a= h^{-1} (a_\constb + \partial_\tau)h$ which is connected to the constant solution $a_\constb$ by gauge transformation $h$:
\begin{equation}
	\partial_\tau \phi + [a, \phi]=0	
\end{equation}
It is also easy to show that the solution is given by
\begin{equation}
	\phi= h^{-1} \varphi h= h^{-1} e^{-a_\constb \tau} c e^{a_\constb \tau} h \label{eq: dilaton solution nonconstant}
\end{equation}
where $c$ is again a constant $sl(N)$ matrix. Note that the solution of the dilaton can be thought as infinitesimal isometry of AdS$_2$, and this can be easily seen from the equation of motion for the dilaton.

The dilaton solutions (or, infinitesimal isometries) are related by similarity transformation if and only if they have identical Casimirs $\mathcal{C}_n$. Here, we define $n$\emph{\tiny th} order Casimir of the dilaton solution $\phi$ by
\begin{equation}
	\mathcal{C}_n\equiv-{1\over n} \tr (\phi^n)
\end{equation}
On the other hand, for a given dilaton solution $\phi$ in~\eqref{eq: dilaton solution nonconstant}, one can find upper-triangular matrix $B$ such that~\cite{Hijano:2013fja}
\begin{equation}
	B^{-1} c B= \begin{pmatrix}
	0 & u_2 & u_3 & \cdots &u_{N-1} & u_N \\
	1 & 0 & 0 &\cdots &0 & 0\\ 
	0 & 1 & 0 & \cdots & 0 & 0 \\
	\vdots &  &  & \cdots & \vdots & \vdots\\
	0 & 0 & 0 & \cdots & 0 & 0\\ 	
	0 & 0 & 0 & \cdots & 1 & 0\\ 
	\end{pmatrix}\equiv K\label{def: matrix k}
\end{equation}
where $u_i$'s (or, the matrix $K$) are defined via
\begin{equation}
	\tr \phi^n=\tr c^n= \tr K^n\hspace{5mm} (n=2,3,\cdots, N)
\end{equation}
Here, by rescaling $B$, one can consider $B$ as constant $SL(N)$ matrix. Hence, for a given connection $a$, two dilaton solutions with identical Casimirs can be related to the same matrix $K$ by similarity transformation, and therefore they are also related by similarity transformation. \ie
\begin{equation}
	\phi'=v^{-1} \phi v\hspace{5mm} \mbox{where} \quad v=h^{-1} e^{-a_\constb \tau} \; c'\; e^{a_\constb \tau} h
\end{equation}
where $c'$ is a constant $SL(N)$ matrix. 
Now, for a given $u_j$ ($j=2,3,\cdots, N$) in~\eqref{def: matrix k}, or equivalently for given eigenvalues $\mu_j$ of the matrix $K$ ($j=1,2,\cdots, N$), we will consider a constant dilaton solution in the $a_\constb$ background. Such a constant dilaton solution should commute with $a_\constb$ by the equation of motion:
\begin{equation}
	[a_\constb, \psi]=0
\end{equation}
Therefore, one can simultaneously diagonalize $a_\constb$ and $\psi$. One can show that $\psi$ is a polynomial of $a_\constb$ of which coefficients are made of Casimirs and the eigenvalues of $a_\constb$ which corresponds to the holonomy. Since $\psi$ is $N\times N$ matrix, it is enough to consider the following polynomial.
\begin{equation}
	\psi= \sum_{j=0}^{N-1} q_j a_\constb^n\qquad\mbox{and}\qquad \tr[\psi]=0\label{eq: constant dilaton ansatz}
\end{equation}
and, we will determine the coefficient $q_j$'s. Since one can simultaneously diagonalize $\psi$ and $a_\constb$, \eqref{eq: constant dilaton ansatz} can be written as follow.
\begin{equation}
	\begin{pmatrix}
	q_0 & q_1 & \cdots & q_{N-1}
	\end{pmatrix}\begin{pmatrix}
	1 & 1 & \cdots & 1\\
	\lambda_1 & \lambda_2 & \cdots & \lambda_N\\
	\lambda_1^2 & \lambda_2^2 & \cdots & \lambda_N^2\\
	\vdots & \vdots & & \vdots\\
	\lambda_1^{N-1} & \lambda_2^{N-1} & \cdots & \lambda_N^{N-1}\\	
	\end{pmatrix}=	\begin{pmatrix}
	\mu_1 & \mu_2 & \cdots & \mu_N
	\end{pmatrix}
\end{equation}
where $\lambda_j$'s are eigenvalues of $a_\constb$ and $\mu_j$'s are the given eigenvalue of $\psi$. By inverting the matrix $(M)_{ij}\equiv \lambda_i^j$, one can determine $q_j$'s in terms of $\lambda$'s and $\mu$'s.

For a general dilaton solution in a non-constant connection $a$, the asymptotic AdS solution that we are interested in is written as
\begin{align}
    a=&h^{-1}(a_\const+ \partial_\tau) h \label{eq: aads2 a}\\
    \phi(\tau) =& h^{-1}(\tau)v^{-1}(\tau)\;\psi\; v (\tau)h(\tau)\label{eq: aads2 phi}
\end{align}
where $v$ is the isometry of the constant background $a_\constb$ in~\eqref{eq: dilaton solution isometry}. Then, using the above argument, one can express $\psi$ in terms of $a_\constb$ as follow.
\begin{equation}
    \psi= \sum_{j=0}^{N-1} q_j a_\constb^j
\end{equation}
where $q_j$ is a constant depending on Casimir of the dilaton and holonomy of the connection. Hence, we have
\begin{equation}
    \phi(\tau)=\sum_{j=0}^{N-1} q_j \left[a - \tilde{h}^{-1} \partial_\tau \tilde{h} \right]^j\label{eq: relation phi and a}
\end{equation}
where we defined $\tilde{h}\equiv vh$. For these solutions, we need to add appropriate boundary condition to lead a consistent variational principle with suitable boundary condition. Unlike AdS$_3$, it is not easy to find such a boundary term and boundary condition which is consistent with \eqref{eq: aads2 a}, \eqref{eq: aads2 phi} and \eqref{eq: relation phi and a}. If there exists such a boundary term and a boundary condition, the reasonable guess for the on-shell action might be written as
\begin{equation}
    S_{\ons}\stackrel{?}{=}  \int d\tau \; \mathcal{C}_2 \mathcal{W}^{(2)}(\tau) 
\end{equation}
where $\mathcal{C}_s$ is $s$\emph{\tiny th} order Casimir. If we consider a dilaton solution of which $\mathcal{C}_2$ is the only non-zero Casimir. \ie $C_s=0$ for ($s>2$). For such a solution, the on-shell action can be written as
\begin{equation}
    S_{\ons}\stackrel{?}{=}  \int d\tau \; \sum_{s=2}^N \mathcal{C}_s \mathcal{W}^{(s)}(\tau) 
\end{equation}
This would be too naive guess. However, it might be to guess that the quadratic on-shell action would be
\begin{equation}
    S_{\ons}= c\sum_{s=2}^N \prod_{j=0}^{s-1}(n^2-j^2) \zeta_{-n}^{(s)} \zeta_n^{(s)}
\end{equation}
Here, we assume that the background has $sl(N)$ isometry, and we demand that the on-shell action vanishes for the $sl(N)$ isometry. Then, the two point function of spin-$s$ soft mode can be read off as

\begin{equation}
    \langle \zeta_n^{(s)}\zeta_{-n}^{(s)} \rangle\sim {1\over c} {1\over \prod_{j=0}^{s-1} (n^2-j^2) }
\end{equation}

To evaluate the OTOC in 2D higher spin gravity, one can consider a Wilson line\footnote{See~\cite{Blommaert:2018oro,Blommaert:2018iqz,Blommaert:2019hjr} for OTOC calculation from Wilson line for Schwarzian theory.} between $(r_1,\tau_1)$ and $(r_2,\tau_2)$:
\begin{align}
    W (r_1,\tau_1;r_2,\tau_2) \equiv \mathcal{P} \exp \left[ -\int^{(r_1,\tau_1)}_{(r_2,\tau_2)}A \right]=b^{-1}(r) e^{-a_\constb (\tau_1-\tau_2)} b(r_2)
\end{align}
And, we define our bi-local field by a particular component of Wilson line:
\begin{align}
    \Phi (\tau_1,\tau_2) \equiv \lim_{r\rightarrow \infty} [e^{2hr}W(r,\tau_1;r,\tau_2)]_{N,1}
\end{align}
where the conformal dimension $h$ is given by $h=-{1\over 2}(N-1)$ like AdS$_3$ case. In the non-constant background in \eqref{eq: aads2 a}, the gravitationally dressed bi-local field can be written as 
\begin{align}
    \Phi^{\text{\tiny dressed}} (\tau_1,\tau_2) \equiv& \lim_{r\rightarrow \infty} \left[e^{2hr}\mathcal{P} \exp \left( -\int^{(r_1,\tau_1)}_{(r_2,\tau_2)}A \right)\right]_{N,1}\cr
    =&\lim_{r\rightarrow \infty}  \left[e^{2hr}b^{-1}(r)h^{-1}(\tau_1)e^{-a_\constb(\tau_1-\tau_2)}h(\tau_2) b(r)\right]_{N,1}
\end{align}
Then, as in AdS$_3$ case, the soft mode expansion of the dressed bi-local field allows us to evaluate OTOCs, which is exactly the same as the OTOTC calculation in Appendix~\ref{app: lyapunov exponent}.

\section{Higher Spin Schwarzian and connections to Toda Theory}
\label{app: Toda Theory}

In Section~\ref{sec: asymptotic ads solution} we only worked to quadratic order in fluctuations. In this appendix we will work out the results for finite fluctuations. For technical reasons we were able to carry out the calculations only at zero temperature and vanishing higher spin charges. The calculation which follows is based on \cite{Li:2015osa}, but such a form has been observed for a long time in various literature~\cite{Bershadsky:1989mf,Marshakov:1989ca,Gonzalez:2018enk,Ma:2019gxy}.

Parametrizing the gauge connection $a_z$ by
\begin{equation}
a_z(z) = L_1 + {1 \over 4} {\cal T}_{(2)}(z) L_{-1} -{1 \over 4} {\cal T}_{(3)}(z) W_{-2}
\end{equation}
To compare with the main text, we have to use ${\cal T}_{(2)} = -{8 \pi \over \kappa_{cs}} {\cal L} ,{\cal T}_{(3)} = -{2 \pi \over \kappa_{cs}} {\mathcal W}$.
The transformation is given by 
\begin{equation}\label{eq:Gauge Transformation}
a_z(z) \rightarrow \tilde a(z) \equiv g(z)^{-1} \bigg( \partial_z + a(z) \bigg) \gamma(z)  = L_1 +  {1 \over 4} \tilde {\cal T}_{(2)}(z) L_{-1} -{1 \over 4} \tilde {\cal T}_{(3)}(z) W_{-2}
\end{equation}
As mentioned in the beginning, we were able to calculate the finite transformations only for the cases when all the initial charges vanish, i.e ${\cal T}_{(2)} = {\cal T}_{(3)} = 0$, i.e $a_z(z) = L_1$. Notice that $\tilde {\cal T}_{(2)}(z) , \tilde {\cal T}_{(3)}(z)$ we read off from \eqref{eq:Gauge Transformation} is the analogue of the schwarzians for the $sl(3)$ case. First note that we can simplify the equations somewhat if we define  $g(z) = e^{-z L_1} \tilde g(z)$. The equation we need to solve for then becomes
\begin{equation}\label{eq:Gauge Transformation Simpler} 
    \tilde g^{-1}(z) \partial_z \tilde g(z)  = L_1 + {1 \over 4} \tilde {\cal T}_{(2)}(z) L_{-1} -{1 \over 4} \tilde {\cal T}_{(3)}(z) W_{-2}
\end{equation}

We will decompose  (the normalizations for the fields are chosen so that the equations turn out to be simple)
\begin{equation}\label{eq:Gauge Parametrization}
\tilde g(z) = e^{ { 1 \over 4} w_2(z) W_2  - \sqrt{2} e_+(z) E_+ - \sqrt{2} f_+(z) F_+ } \  e^{-\phi_1(z) H_1 - \phi_2(z) H_2} \  e^{    {1 \over 4}w_{-2}(z) W_{-2} - { e_-(z) E_- \over \sqrt{2} } - { f_-(z) F_- \over \sqrt 2}   }
\end{equation}
where the $sl(3)$ generators are chosen in chevalley basis
\begin{center}
\begin{equation} \begin{split}
H_1 =\left(
\begin{array}{ccc}
 1 & 0 & 0 \\
 0 & -1 & 0 \\
 0 & 0 & 0 \\
\end{array}
\right)
\quad &
H_2 =\left(
\begin{array}{ccc}
 0 & 0 & 0 \\
 0 & 1 & 0 \\
 0 & 0 & -1 \\
\end{array}
\right) 
\quad 
E_+ =\left(
\begin{array}{ccc}
 0 & 0 & 0 \\
 1 & 0 & 0 \\
 0 & 0 & 0 \\
\end{array}
\right) \quad 
F_+=\left(
\begin{array}{ccc}
 0 & 0 & 0 \\
 0 & 0 & 0 \\
 0 & 1 & 0 \\
\end{array}
\right) \\
& E_-=\left(
\begin{array}{ccc}
 0 & 1 & 0 \\
 0 & 0 & 0 \\
 0 & 0 & 0 \\
\end{array}
\right)\quad F_-=\left(
\begin{array}{ccc}
 0 & 0 & 0 \\
 0 & 0 & 1 \\
 0 & 0 & 0 \\
\end{array}
\right)
\end{split} \end{equation}
\end{center}
We will not need the explicit relation between the Chevally basis and standard basis given in \eqref{eq:sl3 L generators}$\sim$\eqref{eq:sl3 W generators}. Demanding \eqref{eq:Gauge Transformation Simpler} gives 
\begin{alignat}{4}
 e_+' =& -  e^{2 \phi_1 - \phi_2} \;\;,   &e_- =&\phi_1'\\  
 f_+' =& -  e^{2 \phi_2 - \phi_1}\;\;,   &f_- =& \phi_2'\\
 w_2' =& f_+ \partial_+ e_+' - e_+ \partial_+ f_+ \;\;,  \hspace{15mm}  & 4 w_{-2}  = & f_-' -e_-'+ (\phi_1')^2 -  (\phi_2')^2
\end{alignat}
We can solve for $f_\pm , e_\pm , w_\pm$ in terms of $\phi_1,\phi_2$ and finally obtain 
\begin{equation}
\begin{split}
\tilde {\cal T}_{2} & = \left\lbrace \phi_1'' - (\phi_1')^2 + {1 \over 2 } \phi_1' \phi_2' \right\rbrace + \phi_1 \leftrightarrow \phi_2 \\  
& = \left\lbrace  {e_+''' \over e_+'} - {4 \over 3} \left( {e_+'' \over e_+'}\right)^2 - {1 \over 6}\ {e_+'' \over e_+'}\ {f_+'' \over f_+'}  \right\rbrace + e_+ \leftrightarrow f_+\\
2\tilde {\cal T}_{3} & = \left\lbrace  {1 \over 2} \phi_1''' - {1 \over 2} \phi_1'' \phi_2' - \phi_1'' \phi_1'+ \phi_1'^2 \phi_2'\right\rbrace - \phi_1 \leftrightarrow \phi_2 \\
& =  \left\lbrace {1 \over 6} \left[ {e_+'''' \over e_+'} - 5{e_+''' \over e_+'}{e_+'' \over e_+'} + {40 \over 9} \left({e_+'' \over e_+'}\right)^3\right] - {1 \over 6}  {e_+''' \over e_+'}{f_+'' \over f_+'} + {5 \over 18 } {e_+'' \over e_+'}\left({f_+'' \over f_+'}\right)^2 \right\rbrace - e_+ \leftrightarrow f_+
\end{split}
\end{equation}

It is obvious that for the case of pure diffeomorphism we must set $e_+ = f_+$ or equivalently $\phi_1 = \phi_2$. At this point, we can connect the above discussion to Toda field theory. The expression for ${\cal T}_{(2)}, {\cal T}_{(3)}$ in terms of $\phi_1,\phi_2$ are exactly same as that charges of Toda field theory (see eq.~(3.47) of~\cite{Li:2015osa}).


\bibliographystyle{JHEP}
\bibliography{higherspin}

\end{document}